\documentclass[graybox]{svmult}
\usepackage{amscd}

\usepackage{amsmath}
\usepackage{amssymb}

\newtheorem{teo}{Theorem}
\newtheorem{lem}{Lemma}
\newtheorem{pro}{Proposition}
\newtheorem{cor}{Corollary}

\newtheorem{defi}{Definition}

\newtheorem{examp}{Example}

\def\sp{\hspace{0.2cm}}

\def\qee{\begin{flushright} $\Diamond$ \end{flushright}}
\def\ov{\overline}

\begin{document}

\title*{A dynamical point of view of Quantum Information: entropy and pressure}
\titlerunning{A dynamical point of view of Quantum Information}

\author{A. Baraviera, C. F. Lardizabal,  A. O. Lopes, and M. Terra Cunha}

\institute{A. T. Baraviera \at I.M. - UFRGS, Porto Alegre - 91500-000, Brasil, \email{atbaraviera@gmail.com}
\and C. F. Lardizabal \at I.M. - UFRGS, Porto Alegre - 91500-000, Brasil, \email{carlos.lardizabal@gmail.com }
\and A. O. Lopes \at I.M. - UFRGS, Porto Alegre - 91500-000, Brasil, \email{arturoscar.lopes@gmail.com}
\and M. Terra Cunha  \at D. M - UFMG, Belo Horizonte - 30161-970, Brasil, \email{marcelo.terra.cunha@gmail.com}
}

\maketitle

\abstract{Quantum Information is a new area of research which
has been growing rapidly since last decade. This topic is very  close to
potential applications to the so called Quantum Computer. In our
point of view it makes sense to develop a more ``dynamical point of
view" of this theory. We want to consider the concepts of entropy
and pressure for ``stationary systems" acting on density matrices
which generalize the usual ones in Ergodic Theory (in the sense of
the Thermodynamic Formalism of R. Bowen, Y. Sinai and D. Ruelle).
We consider the operator $\mathcal{L}$ acting on density matrices
$\rho\in \mathcal{M}_N$ over a finite $N$-dimensional complex
Hilbert space
$\mathcal{L}(\rho):=\sum_{i=1}^k tr(W_i\rho W_i^*)V_i\rho V_i^*,$ where $W_i$ and $V_i$,
$i=1,2,\dots k$ are operators in this Hilbert space.
$\mathcal{L}$ is not a linear operator. In some sense this
operator is a version of  an Iterated Function System (IFS).
Namely, the $V_i\,(.)\,V_i^*=:F_i(.)$, $i=1,2,\dots,k$, play the
role of the inverse branches (acting on the configuration space of
density matrices $\rho$) and the $W_i$ play the role of the
weights one can consider on the IFS. We suppose that for all
$\rho$ we have that $\sum_{i=1}^k tr(W_i\rho W_i^*)=1$.  A family
$W:=\{W_i\}_{i=1,\dots, k}$ determines a Quantum Iterated Function
System (QIFS) $\mathcal{F}_{W}$,
$\mathcal{F}_W=\{\mathcal{M}_N,F_i,W_i\}_{i=1,\dots, k}.$}

Dynamics, Games and Science I, DYNA 2008, Edit. M. Peixoto, A. Pinto and D Rand, pp 81-122 Springer Verlag (2011)

\section{Introduction}

We will present a survey, and also some new results, of certain
topics in Quantum Information from a strictly mathematical point
of view. This area is very close to potential applications to the
so called Quantum Computer \cite{R1}. In our point of view it makes sense to develop
a more ``dynamical point of view" of this theory. For instance,
Von Neumann entropy is a very nice and useful concept, but, in our
point of view, it is not a dynamical entropy. A nice exposition
about this theory from an Ergodic Theory point of view is
presented in \cite{benatti} (see also \cite{bcs}). Our  setting is different. Part of
our work is to justify why the concepts we present here are
natural generalizations of the usual ones in Thermodynamic
Formalism.

We have to analyze first the fundamental concepts in both
theories. It is well-known that the so called Quantum Stochastic
Processes have some  special features which present a quite
different nature than  the usual classical Stochastic Processes. A
main issue on QSP is the possibility of interference (see
\cite{BLLT} \cite{BLLT2} \cite{GZ} \cite{slom} \cite{Sr}). We will
analyze carefully Quantum Iterated Function Systems, which were
described previously by \cite {lozinski} and \cite{wsbook}.

We refer the reader to \cite{BLLT} for the proofs of the results
presented in  the first part of this exposition.

Density matrices play the role of probabilities on Quantum Mechanics. In this work we
investigate a generalization of the classical Thermodynamic
Formalism (in the sense of Bowen, Sinai and Ruelle) for the
setting of density matrices. We consider the operator
$\mathcal{L}$ acting on density matrices $\rho\in \mathcal{M}_N$
over a finite $N$-dimensional complex Hilbert space
$$
\mathcal{L}(\rho):=\sum_{i=1}^k tr(W_i\rho W_i^*)V_i\rho V_i^*,
$$
where $W_i$ and $V_i$, $i=1,2,..k$ are operators in this Hilbert
space. Note that $\mathcal{L}$ is not a linear operator.

In some sense this operator is a version of  an Iterated Function System (IFS). Namely, the $V_i\,(.)\,V_i^*=:F_i(.)$, $i=1,2,\dots,k$, play the role of the inverse branches (acting on the configuration space of density matrices $\rho$) and the $W_i$ play the role of the weights one can consider on the IFS. We suppose that for all $\rho$ we have that $\sum_{i=1}^k tr(W_i\rho W_i^*)=1$. This means that $ \mathcal{L}_{\mathcal{F}_W}$ is a normalized operator.

A  family $W:=\{W_i\}_{i=1,\dots, k}$ determines a Quantum Iterated Function System (QIFS) $\mathcal{F}_{W}$,
$$\mathcal{F}_W=\{\mathcal{M}_N,F_i,W_i\}_{i=1,\dots, k}$$

We want to consider a new concept of entropy for stationary systems
acting on density matrices which generalizes the usual one in
Ergodic Theory. In our setting the $V_i$, $i=1,2,\dots,k$ are fixed
(i.e. the dynamics of the inverse branches is fixed in the beginning) and we
consider the different families $W_i$, $i=1,2,\dots,k$, (also with
the attached corresponding eigendensity matrix $\rho_W$) as possible
Jacobians (of ``stationary probabilities").

It is appropriate to make here a remark about the meaning of ``stationarity" for us.
In Ergodic Theory the action of the shift $\sigma$ in the
Bernoulli space $\Omega=\{1,2,\dots,k\}^\mathbb{N}$  with $k$ symbols
is well understood. The concept of stationarity for a Stochastic
Process (where the space of states is $S=  \{1,2,\dots,k \}$) is
defined by the shift-invariance for the
associated probability $P$ in the Bernoulli space (the space of
paths). Shannon-Kolmogorov entropy is a concept designed for
stationary probabilities. When the probability $P$ is associated
to a Markov chain, this entropy is given by
$$
H(P):=-\sum_{i,j=1}^Np_i p_{ij}\log{p_{ij}} ,$$ where $P=(p_{ij})$
describes the transition matrix, and $p_i$ the invariant
probability vector, $i,j= 1,2,..,k$. This is the key idea for our
definition of stationary entropy.

Thermodynamic Formalism and the Ruelle operator for a potential
$A:\Omega \to \mathbb{R}$ are natural generalizations of the
theory associated to the Perron theorem  for positive matrices
(see \cite{Sp}) (this occurs when the potential depends on only
the first two symbols of $w=(w_1,w_2,w_3,\dots)\in \Omega$). We will
analyze the Pressure problem for density matrices under this last
perspective.

The main point here (and also in \cite{BLLT} \cite{BLLT2} \cite{lopes_elismar} \cite{lopes}) is that in order to define Kolmogorov entropy one can avoid the use of partitions, etc. We just need to look the problem at the level of  Ruelle operators (which in some sense captures the underlying dynamics).

Given a normalized family $W_i$, $i=1,2,..,k$, a natural
definition of entropy, denoted by $h_V(W)$, is given by
$$
-\sum_{i=1}^k \frac{tr(W_i\rho_W W_i^*)}{tr(V_i\rho_W V_i^*)}\sum_{j=1}^k tr\Big(W_j V_i\rho_W V_i^*
W_j^*\Big)\log{\Big(\frac{tr(W_j V_i\rho_W V_i^* W_j^*)}{tr(V_i\rho_W V_i^*)}\Big)},
$$
where, $\rho_W$ denotes the barycenter of the unique invariant,
attractive measure for the Markov operator $\mathcal{V}$ associated
to $\mathcal{F}_W$. We show that this generalizes the entropy of a
Markov System. This will be described later on this work.

A different definition of entropy for density operators is presented in \cite{BLLT2} \cite{GL}. There are examples where the values one gets from these two concepts are different (see \cite{BLLT2}).

We also want to present here a concept of pressure for stationary systems acting on density matrices  which generalizes the usual one in Ergodic Theory.

In addition to the dynamics obtained by the $V_i$, which are fixed, a family of potentials $H_i$,  $i=1,2,\dots k$ induces a kind of Ruelle operator given by
\begin{equation}\label{operador_usa1a}
\mathcal{L}_H(\rho):=\sum_{i=1}^k tr(H_i\rho H_i^*)V_i\rho V_i^*
\end{equation}

We show that such operator admits an eigenvalue $\beta$ and an
associated eigenstate $\rho_{\beta} $, that is, one satisfying
$\mathcal{L}_H(\rho_{\beta})=\beta \rho_{\beta}$.

The natural generalization of the concept of pressure for a family
$H_i$, $i=1,2,\dots k$ is the problem of finding the maximization on
the possible normalized families $W_i$, $i=1,2,\dots k$, of the
expression
$$
h_V(W)+ \sum_{j=1}^k \log \Big( tr(H_j\rho_{H}
H_j^*)tr(V_j\rho_{H} V_j^*)\Big) tr(W_j \rho_W  W_j^*)$$ We show
a relation between the eigendensity matrix $\rho_H$ for the Ruelle
operator and the set of $W_i$, $i=1,2,\dots k$, which maximizes
pressure. In the case each $V_i$, $i=1,2,\dots k$, is unitary, then
the maximum value is $\log \beta$.

Our work is  inspired by the results presented in \cite {lozinski}
and \cite{wsbook}. We would like to thank these authors for
supplying us with the corresponding references.

We point out that completely positive mappings (operators) acting on density matrices are of great importance in Quantum Computing. These operators can be written  in the Stinespring-Kraus form. This motivates the study of operators in the class we will assume here, which are a generalization of such Stinespring-Kraus transformations.

The initial part of our work is dedicated  to present all the
definitions and concepts that are not well-known (at least for the
general audience of people in Dynamical Systems), in a systematic
and well organized way. We present many examples and all the basic
main definitions which are necessary to understand the theory.
However, we do not have the intention to exhaust what is already
known. We believe that the theoretical results presented here can
be useful as a general tool to understand problems in Quantum
Computing.

Several examples are presented with all details in the text.  We believe  that this
will help the reader to understand the main issues of the theory.

In order to simplify the notation we will present most of our results for the case of two by two matrices.

In sections \ref{sec_basic} and \ref{sec_examples} we present some basic definitions, examples and we show some preliminary relations of our setting to the classical Thermodynamic Formalism. In section \ref{sec_eigenvalues} we present an eigenvalue problem for non-normalized Ruelle operators  which will be required later. Some properties and concepts about density matrices and Ruelle operators are presented in sections \ref{sec_densities} and \ref{sec_somelemmas}. In section \ref{sec_novaentr} we introduce the concept of stationary entropy for {\it measures} defined on the set of density matrices. In section \ref{camarkov} we compare this definition with the usual one for Markov Chains.
Section \ref{sec_cc} aims to motivate the interest on pressure and the capacity-cost function.  The sections \ref{sec_analysis1}, \ref{sec_analysis2}, \ref{sec_analysis3} and \ref{sec_analysis4} are dedicated to the presentation of our main results on pressure, important inequalities, examples and its relation with the classical theory of Thermodynamic Formalism.

This work is part of the thesis dissertation of C. F. Lardizabal in Prog. Pos-Grad. Mat. UFRGS (Brazil) \cite{lar}.

\section{Basic definitions}\label{sec_basic}

Let $M_N(\mathbb{C})$ the set of complex matrices  of order $N$. If $\rho\in M_N(\mathbb{C})$ then $\rho^*$ denotes the transpose conjugate of $\rho$. We consider in $\mathbb{C}^N$ the ${\cal L}^2$ norm. A state (or vector) in $\mathbb{C}^N$ will be denoted by $\psi$ or $\vert \psi\rangle$, and the associated projection will be written $\vert \psi\rangle\langle\psi\vert$. Define
$$\mathcal{H}_N:=\{\rho\in M_N(\mathbb{C}):\rho^*=\rho\}$$
$$\mathcal{PH}_N:=\{\rho\in \mathcal{H}_N:\langle\rho \psi,\psi\rangle\geq 0 , \forall \psi\in\mathbb{C}^N\}$$
$$\mathcal{M}_N:=\{\rho\in \mathcal{PH}_N: tr(\rho)=1\}$$
$$\mathcal{P}_N:=\{\rho\in \mathcal{H}_N: \rho=\vert\psi\rangle\langle\psi\vert,\psi\in\mathbb{C}^N, \langle\psi\vert\psi\rangle=1\},$$
the space of hermitian, positive, density operators and pure states, respectively. Density operators are also called mixed states. Any state $\rho$, by the spectral theorem,  can be written as
\begin{equation}
\rho=\sum_{i=1}^k p_i\vert \psi_i\rangle\langle \psi_i\vert,
\end{equation}
 for some choice of $p_i$, which are positive numbers with $\sum_i p_i=1$, and $\psi_i$, which have norm one and are orthogonal.

The set $\mathcal{P}_N$ is the set of extremal points of
$\mathcal{M}_N$, that is, the set of  points which can not be decomposed as a nontrivial convex combination of elements in $\mathcal{M}_N.$

\begin{defi}
Let $G_i:\mathcal{M}_N\to\mathcal{M}_N$, $p_i:\mathcal{M}_N\to[0,1]$, $i=1,\dots ,k$ and such that $\sum_i p_i(\rho)=1$. We call
\begin{equation}\label{qifs}
\mathcal{F}_N=\{\mathcal{M}_N, G_i, p_i:i=1,\dots, k\}
\end{equation}
a {\bf Quantum Iterated Function System} (QIFS).
\end{defi}

\begin{defi}
A QIFS is {\bf homogeneous} if $p_i$ and $G_ip_i$ are affine mappings, $i=1,\dots , k$.
\end{defi}

Suppose that the QIFS considered is such that there are $V_i$ and $W_i$ linear maps, $i=1,\dots, k$, with $\sum_{i=1}^k W_i^*W_i=I$ such that
\begin{equation}
G_i(\rho)=\frac{V_i\rho V_i^*}{tr(V_i\rho V_i^*)}
\end{equation}
and
\begin{equation}
p_i(\rho)=tr(W_i\rho W_i^*)
\end{equation}
Then we have that a QIFS is homogeneous if $V_i$=$W_i$, $i=1,\dots,k$.

Now we can define a Markov operator
$\mathcal{V}:\mathcal{M}(\mathcal{M}_N)\to\mathcal{M}(\mathcal{M}_N)$,
$$(\mathcal{V}\mu)(B)=\sum_{i=1}^k\int_{G_i^{-1}(B)}p_i(\rho)d\mu(\rho),$$
where $\mathcal{M}(\mathcal{M}_N)$ denotes the space of probability measure over $\mathcal{M}_N$. We also define $\Lambda: \mathcal{M}_N\to\mathcal{M}_N$,
$$\Lambda(\rho):=\sum_{i=1}^k p_i(\rho)G_i(\rho)$$

The operator defined above has no counterpart in the classical Thermodynamic Formalism. We will also consider the operator acting on density matrices $\rho$.

$$\mathcal{L}(\rho)=\sum_{i=k}^k q_i(\rho)V_i\rho V_i^{*}.$$

If for all $\rho$ we have $\sum_{i=k}^k q_i(\rho)=1$, we say the operator is normalized.

In the normalized case, the different possible choices of $q_i, i=1,2,\dots,k$, (which means different choices of $W_i, i=1,2,\dots,k$) play here the role of the different Jacobians  of possible invariant probabilities (see \cite{Man} II. 1, and  \cite{lopes}) in Thermodynamic Formalism. In some sense the  probabilites can be identified with the Jacobians (this is true at least for Gibbs probabilities of H\"older potentials \cite{par}). The set of Gibbs probabilities for H\"older potentials is dense in the set of invariant probabilities \cite{lop1}.

 We are also interested on the non-normalized case.
If the QIFS is homogeneous, then
\begin{equation}\label{lambda_um}
\Lambda(\rho)=\sum_i V_i\rho V_i^*
\end{equation}

\begin{teo}\cite{wsbook} A mixed state $\rho_0$ is $\Lambda$-invariant if and only if
\begin{equation}\label{bari}
\rho_0=\int_{\mathcal{M}_N} \rho d\mu(\rho),
\end{equation}
for some $\mathcal{V}$-invariant measure $\mu$.
\end{teo}

\bigskip

In order to define hyperbolic QIFS, one has to define a distance on the space of mixed states. For instance, we could choose one of the following:
$$D(\rho_1,\rho_2)=\sqrt{tr[(\rho_1-\rho_2)^2]}$$
$$D(\rho_1,\rho_2)=tr\sqrt{(\rho_1-\rho_2)^2}$$
$$D(\rho_1,\rho_2)=\sqrt{2\{1-tr[(\rho_1^{1/2}\rho_2\rho_1^{1/2})^{1/2}]\}}$$
Such metrics generate the same topology on $\mathcal{M}$. Considering the space of mixed states with one of those metrics we can make the following definition. We say that a QIFS is {\bf hyperbolic} if the quantum maps $G_i$ are contractions with respect to one of the distances on $\mathcal{M}_N$ and if the maps $p_i$ are H\"older-continuous and positive, see for instance, \cite{lozinski}.

\bigskip

\begin{pro} If a QIFS (\ref{qifs}) is homogeneous and hyperbolic the associated Markov operator admits a unique invariant measure $\mu$. Such invariant measure determines a unique  $\Lambda$-invariant state $\rho\in\mathcal{M}_N$, given by (\ref{bari}).
\end{pro}

See \cite{lozinski}, \cite{wsbook} for the proof.

\section{Examples of QIFS}\label{sec_examples}

\begin{examp}\label{exemp7} $\Omega=\mathcal{M}_N$, $k=2$, $p_1=p_2=1/2$, $G_1(\rho)=U_1\rho U_1^*$, $G_2(\rho)=U_2\rho U_2^*$. The normalized identity matrix $\rho_*=I/N$ is $\Lambda$-invariant, for any choice of unitary $U_1$ and $U_2$. Note that we can write
$$\rho_*=\int_{\mathcal{M}_N}\rho d\mu(\rho)$$
where the measure $\mu$, uniformly distributed over
$\mathcal{P}_N$, is $\mathcal{V}$-invariant.
\end{examp}

\qee

In the example described below we use Dirac notation for the
projections.

\begin{examp}

We are interested in finding the fixed point $\hat{\rho}$  for $\Lambda$ in an example for the case $N=2$ and $k=3$.

Consider the bits $|0>=(0,1)$ and $|1>=(1,0)$ (the canonical basis). The states $\rho$ are generated by $|0><0|$, $|0><1|$, $|1><0|$ and $|1><1|$. Take $V_1 =I$ and $V_2$ such that  $|0>\, \to\, |0>$ and $|1>\, \to\, |0>$. Consider $ V_3$  such that  $|0>\, \to\, |1>$ and $|1>\, \to\, |1>$. That is, $ V_2= |0><0| \,+\, |0><1|$ and $  V_3= |1><0| \,+\, |1><1|$. Therefore,
$ V_2^*= |0><0| \,+\, |1><0|$ and $  V_3^*= |0><1| \,+\, |1><1|$.
Suppose $p_i=\hat{p}_i$, $i=1,2,3$,  are such that
$\sum_i p_i \,\,=\,1$ (in this case,  each $p_i$ is independent of   $\rho$). Therefore, we consider the operator $\mathcal{L}$ and  look for fixed points $\rho$.
Suppose
$$\rho=\rho_{00}\, |0><0|\,+ \, \rho_{01} \,|0><1|\, +  \rho_{10}\,|1><0|\,+\, \rho_{11}\,|1><1|$$
Then
$$
   \Lambda(\rho) = \sum_{i=1}^3\, p_i(\rho) \,    \frac{( V_i\,\rho\, V_i^*)}{ \text{tr}\, ( V_i\,\rho\, V_i^*)} =
$$
$$ \sum_{i=1}^3 p_i \,[\frac{ \,V_i \,(\,  \,(\rho_{00}\, |0><0|\,+ \, \rho_{01} \,|0><1|\, +  \rho_{10}\,|1><0|\,+\, \rho_{11}\,|1><1|   \,\,  ) \,)\, V_i^*}{  \text{tr}\, ( V_i\,\rho\, V_i^*)} ]
$$

Let us compute first the action of the operator $V_2  |0><0| V_2^*$.

Note that $(\,V_2  |0><0| V_2^*\,)\,|0>=   V_2  |0><0| \, (\,|0>+|1>\,) = V_2 |0>= |0>$ and
$(\,V_2  |0><0| V_2^*\,)\, |1>=   V_2  (0)= 0$. More generally
$$\rho \,V_2^* =  \,\,\, (          \,\, \rho_{00}\, |0><0|\,+$$
 $$\, \rho_{01} \,|0><1|\, +  \rho_{10}\,|1><0|\,+\, \rho_{11}\,|1><1|  \,\,)\,\, (  |0><0| \,+\, |1><0| )=$$
$$        \,\, \rho_{00}\, |0><0| +         \rho_{01} \,|0><0|\, +\,  \rho_{10}\, |1><0| +         \rho_{11} \,|1><0|\,.$$

Therefore,
$$V_2 \, \rho \,V_2^* = (|0><0| \,+\, |0><1|)\,( \,\, \rho_{00}\, |0><0| +$$
$$         \rho_{01} \,|0><0|\, +\,  \rho_{10}\, |1><0| +         \rho_{11} \,|1><0|\,)=$$
$$   (\,\, \rho_{00}\, +         \rho_{01} \,\, +\,  \rho_{10}\, +         \rho_{11} \,)\,|0><0|=(1+ 2 Re(\,\rho_{01})) \,\, |0><0|, $$
because $\rho$ has trace $1= \rho_{00}+ \rho_{11}$. Note that $tr(V_2 \, \rho \,V_2^*)=(1+ 2 Re(\,\rho_{01})).$ A similar result can be obtained  for $V_3$. Proceeding in the same way we get that
$$   \Lambda(\rho) = \,p_1 \,(  \,\, \rho_{00}\,\, |0><0|\,+ \, \rho_{01}\, \,|0><1|\, +  \rho_{10}\,\,|1><0|\,+\, \rho_{11}\,\,|1><1|  \,\,  )\, +$$
$$ p_2 \,\,|0><0|+ p_3\,\, |1><1|.$$
The equation
$$ \Lambda( \rho)= \rho= \,\, \rho_{00}\,\, |0><0|\,+ \, \rho_{01}\, \,|0><1|\, +  \rho_{10}\,\,|1><0|+\, \rho_{11}\,\,|1><1|  $$
means
$$ p_1 \, \rho_{00} + p_2\, = \rho_{00},$$
$$ p_1 \, \rho_{01} = \rho_{01},$$
$$ p_1 \, \rho_{10}  = \rho_{10},$$
$$ p_1 \, \rho_{11} + p_3 = \rho_{11}.$$

If $p_1\neq0$, then $\rho_{01}=\rho_{10}=0$. Finally, if $p_1\neq1 $, then $ \rho_{00}= \frac{p_2}{1-p_1}$ and
$ \rho_{11}= \frac{p_3}{1-p_1}$ and the fixed point is
$$\hat{\rho}= \frac{p_2}{1-p_1}\,|0><0|
+ \frac{p_3}{1-p_1}\,|1><1|.$$

\end{examp}

\qee

We recall that a mapping $\Lambda$ is {\bf completely positive} (CP) if $\Lambda\otimes I$ is positive for any extension of the Hilbert space considered $\mathcal{H}_N\to\mathcal{H}_N\otimes\mathcal{H}_E$. We know that every CP mapping which is trace-preserving can be represented (in a nonunique way) in the Stinespring-Kraus form
$$\Lambda_K(\rho)=\sum_{j=1}^k V_j\rho V_j^*, \sp \sum_{j=1}^k V_j^* V_j=1,$$
where the $V_i$ are linear operators. Moreover if we have $\sum_{j=1}^k V_j  V_j^*=I$, then $\Lambda(I/N)=I/N$. This is the case if each of the $V_i$ are normal.

\bigskip

We call a unitary trace-preserving CP map a {\bf bistochastic map}. An example of such a mapping is
$$\Lambda_U(\rho)=\sum_{i=1}^k p_i U_i\rho U_i^*,$$
where the $U_i$ are unitary operators and $\sum_i p_i=1$. Note that if we write $G_i(\rho)=U_i\rho U_i^*$, then example \ref{exemp7} is part of this class of operators. For such operators we have that $\rho_*$ is an invariant state for  $\Lambda_U$ and also that $\delta_{\rho_*}$ is invariant for the Markov operator $P_U$ induced by this QIFS.

\bigskip
We will present a simple example of the kind of problems we are
interested here, namely eigenvalues and eigendensity matrices. Let
$\mathcal{H}_N$ be a Hilbert space of dimension $N$.  As before,
let $\mathcal{M}_N$ be the space of density operators on
$\mathcal{H}_N$. A natural problem is to find fixed points for
$\Lambda:\mathcal{M}_N\to\mathcal{M}_N$,
$$\Lambda(\rho)=\sum_{i=1}^k V_i\rho V_i^*.$$

In order to simplify our reasoning we fix $N=2$ and $k=2$. Let
$$V_1=\left(
\begin{array}{cc}
v_1 & v_2\\
v_3 & v_4
\end{array}
\right),
\sp V_2=\left(
\begin{array}{cc}
w_1 & w_2\\
w_3 & w_4
\end{array}
\right),
\sp \rho=\left(
\begin{array}{cc}
\rho_1 & \rho_2\\
\ov{\rho_2} & \rho_4
\end{array}
\right),
$$
 where $V_1$ and $V_2$ are invertible and $\rho$ is a density operator. We would like to find $\rho$ such that
\begin{equation}\label{equm}
V_1\rho V_1^*+V_2\rho V_2^*=\rho.
\end{equation}

Below we have an example where the matrices $V_i$ are not real.
\begin{examp}\label{vex1} Let
$$V_1=e^{i\,k}\left(
\begin{array}{cc}
\sqrt{p} & 0\\
0 & -\sqrt{p}
\end{array}
\right),
\sp V_2=e^{i\,l}\left(
\begin{array}{cc}
\sqrt{1-p} & 0\\
0 & -\sqrt{1-p}
\end{array}
\right),
$$
where $k,l\in\mathbb{R}$, $p\in (0,1)$. Then $V_1^* V_1+V_2^* V_2=I$. A simple calculation shows that $\rho_2=0$, and then
$$\rho=\left(
\begin{array}{cc}
q & 0\\
0 & 1-q
\end{array}
\right)
$$
is invariant to $\Lambda(\rho)=V_1\rho V_1^*+V_2\rho V_2^*$, for  $q\in (0,1)$.
\end{examp}

\qee

Now we make a few considerations about the Ruelle operator $\mathcal{L}$ defined before. In particular, we show that Perron's classic eigenvalue problem is a particular case of the problem for the operator $\mathcal{L}$ acting on matrices. Let
$$V_1=\left(
\begin{array}{cc}
p_{00} & 0\\
0 & 0
\end{array}
\right),
\sp V_2=\left(
\begin{array}{cc}
0 & p_{01}\\
0 & 0
\end{array}
\right)
$$
$$V_3=\left(
\begin{array}{cc}
0 & 0\\
p_{10} & 0
\end{array}
\right),
\sp V_4=\left(
\begin{array}{cc}
0 & 0\\
0 & p_{11}
\end{array}
\right),
\sp \rho=\left(
\begin{array}{cc}
\rho_1 & \rho_2\\
\rho_3 & \rho_4
\end{array}
\right)
$$
Define
$$\mathcal{L}(\rho)=\sum_{i=1}^4q_i(\rho)V_i\rho V_i^{*}$$
We have that $\mathcal{L}(\rho)=\rho$ implies $\rho_2=0$ and
\begin{equation}\label{densist1}
a\rho_1+b\rho_4=\rho_1
\end{equation}
\begin{equation}\label{densist2}
c\rho_1+d\rho_4=\rho_4
\end{equation}
where
$$a=q_1p_{00}^2,\sp b=q_2p_{01}^2,\sp c=q_3p_{10}^2,\sp d=q_4p_{11}^2$$
Solving (\ref{densist1}) and (\ref{densist2}) in terms of $\rho_1$ gives
$$\rho_1=\frac{b}{1-a}\rho_4,\sp \rho_1=\frac{1-d}{c}\rho_4$$
that is,
\begin{equation}\label{condsq}
\frac{b}{1-a}=\frac{1-d}{c}
\end{equation}
which is a restriction over the $q_i$. For simplicity we assume here that the $q_i$ are constant. One can show that
\begin{equation}\label{pirhoc2}
\rho=\left(
\begin{array}{cc}
\frac{q_2p_{01}^2}{q_2p_{01}^2-q_1p_{00}^2+1} & 0\\
0 & \frac{1-q_1p_{00}^2}{q_2p_{01}^2-q_1p_{00}^2+1}
\end{array}
\right)=\left(
\begin{array}{cc}
\frac{1-q_4p_{11}^2}{1-q_4p_{11}^2+q_3p_{10}^2} & 0\\
0 & \frac{q_3p_{10}^2}{1-q_4p_{11}^2+q_3p_{10}^2}
\end{array}
\right)
\end{equation}
Now let
$$P=\sum_i V_i=\left(
\begin{array}{cc}
p_{00} & p_{01}\\
p_{10} & p_{11}
\end{array}
\right),$$ be a column-stochastic matrix. Let $\pi=(\pi_1,\pi_2)$ such that $P\pi=\pi$. Then

\begin{equation}\label{pirhoc1}
\pi=(\frac{p_{01}}{p_{01}-p_{00}+1},\frac{1-p_{00}}{p_{01}-p_{00}+1})\end{equation}

Comparing (\ref{pirhoc1}) and (\ref{pirhoc2}) suggests that we should fix
\begin{equation}\label{solnat}
q_1=\frac{1}{p_{00}},\sp q_2=\frac{1}{p_{01}},\sp q_3=\frac{1}{p_{10}}, \sp q_4=\frac{1}{p_{11}}
\end{equation}
Then the nonzero entries of $\rho$ are equal to the entries of $\pi$ and therefore we associate the fixed point of $P$ to the fixed point of some $\mathcal{L}$ in a natural way. But note that such a choice of $q_i$ is not unique, because
\begin{equation}\label{sol4mat1}
q_2=\frac{1-q_1p_{00}^2}{p_{01}p_{10}},\sp q_4=\frac{1-q_3p_{10}p_{01}}{p_{11}^2},
\end{equation}
for any $q_1,q_3$ also produces $\rho$ with nonzero coordinates equal to the coordinates of $\pi$.

\bigskip

Now we consider the following problem. Let
$$V_1=\left(
\begin{array}{cc}
h_{00} & 0\\
0 & 0
\end{array}
\right),
\sp V_2=\left(
\begin{array}{cc}
0 & h_{01}\\
0 & 0
\end{array}
\right)
,\sp V_3=\left(
\begin{array}{cc}
0 & 0\\
h_{10} & 0
\end{array}
\right)$$
$$V_4=\left(
\begin{array}{cc}
0 & 0\\
0 & h_{11}
\end{array}
\right),
\sp H=\sum_i V_i,
\sp \rho=\left(
\begin{array}{cc}
\rho_1 & \rho_2\\
\rho_3 & \rho_4
\end{array}
\right)
$$

Define
$$\mathcal{L}(\rho)=\sum_{i=1}^4 q_iV_i\rho V_i^*,$$
where $q_i\in\mathbb{R}$. Assume that $h_{ij}\in\mathbb{R}$, so we want to obtain $\lambda$ such that
$\mathcal{L}(\rho)=\lambda\rho$, $\lambda\neq 0$, and $\lambda$ is the largest eigenvalue. With a few calculations we obtain $\rho_2=\rho_3=0$,
$$q_1h_{00}^2\rho_1+q_2h_{01}^2\rho_4=\lambda\rho_1$$
$$q_3h_{10}^2\rho_1+q_4h_{11}^2\rho_4=\lambda\rho_4$$
that is,
\begin{equation}\label{yetanother1}
a\rho_1+b\rho_4=\lambda\rho_1
\end{equation}
\begin{equation}\label{yetanother2}
c\rho_1+d\rho_4=\lambda\rho_4,
\end{equation}
with
$$a=q_1h_{00}^2,\sp b=q_2h_{01}^2,\sp c=q_3h_{10}^2,\sp d=q_4h_{11}^2$$
Therefore
$$\rho=\left(
\begin{array}{cc}
\frac{\lambda-d}{c}\rho_4 & 0\\
0 & \rho_4
\end{array}
\right)=\left(
\begin{array}{cc}
\frac{b}{\lambda-a}\rho_4 & 0\\
0 & \rho_4
\end{array}
\right)
$$
and
$$\frac{\lambda-d}{c}=\frac{b}{\lambda-a}$$
Solving for $\lambda$, we obtain the eigenvalues
$$\lambda=\frac{a+d}{2}\pm\frac{\zeta}{2}=\frac{a+d}{2}\pm\frac{\sqrt{(d-a)^2+4bc}}{2}$$
$$=\frac{1}{2}\Big(q_1h_{00}^2+q_4h_{11}^2\pm\sqrt{(q_4h_{11}^2-q_1h_{00}^2)^2+4q_2q_3h_{01}^2h_{10}^2}\Big),$$
where
$$\zeta=\sqrt{(d-a)^2+4bc}=\sqrt{(q_4h_{11}^2-q_1h_{00}^2)^2+4q_2q_3h_{01}^2h_{10}^2}$$
and the associated eigenfunctions
$$\rho=\left(
\begin{array}{cc}
\frac{a-d\pm\zeta}{2c}\rho_4 & 0\\
0 & \rho_4
\end{array}
\right)=\left(
\begin{array}{cc}
\frac{2b}{d-a\pm\zeta}\rho_4 & 0\\
0 & \rho_4
\end{array}
\right)
$$
But $\rho_1+\rho_4=1$ so we obtain
$$
\rho=\left(
\begin{array}{cc}
\frac{a-d\pm\zeta}{a-d\pm\zeta+2c} & 0\\
0 & \frac{2c}{a-d\pm\zeta+2c}
\end{array}
\right)$$
\begin{equation}\label{eqcomplvejaa}
=\left(
\begin{array}{cc}
\frac{q_1h_{00}^2-q_4h_{11}^2\pm\zeta}{q_1h_{00}^2-q_4h_{11}^2\pm\zeta+2q_3h_{10}^2} & 0\\
0 & \frac{2q_3h_{10}^2}{q_1h_{00}^2-q_4h_{11}^2\pm\zeta+2q_3h_{10}^2}
\end{array}
\right)
\end{equation}
that is,
$$
\rho=\left(
\begin{array}{cc}
\frac{-2b}{a-2b-d\mp\zeta} & 0\\
0 & \frac{a-d\mp\zeta}{a-2b-d\mp\zeta}
\end{array}
\right)$$
\begin{equation}\label{eqcomplveja}
=\left(
\begin{array}{cc}
\frac{-2q_2h_{01}^2}{q_1h_{00}^2-2q_2h_{01}^2-q_4h_{11}^2\mp\zeta} & 0\\
0 & \frac{q_1h_{00}^2-q_4h_{11}^2\mp\zeta}{q_1h_{00}^2-2q_2h_{01}^2-q_4h_{11}^2\mp\zeta}
\end{array}
\right)
\end{equation}
Therefore we obtained that $\rho_1,\rho_4, q_1,\dots,q_4, \lambda$ are implicit solutions for the set of equations (\ref{yetanother1})-(\ref{yetanother2}). Recall that in this case we obtained $\rho_2=\rho_3=0.$

\bigskip
Now we consider the problem of finding the eigenvector associated to the dominant eigenvalue of $H$. The eigenvalues are
$$\lambda=\frac{1}{2}\Big(h_{00}+h_{11}\pm\sqrt{(h_{00}-h_{11})^2+4h_{01}h_{10}}\Big)$$
Then we can find $v$ such that $Hv = \lambda v$ from the set of equations
\begin{equation}\label{yetanother3}
h_{00}v_1+h_{01}v_2=\lambda v_1
\end{equation}
\begin{equation}\label{yetanother4}
h_{10}v_1+h_{11}v_2=\lambda v_2
\end{equation}
which determine $v_1, v_2,\lambda$ implicitly. Note that if we set

$$q_1=\frac{1}{p_{00}},\sp q_2=\frac{1}{p_{01}},\sp q_3=\frac{1}{p_{10}}, \sp q_4=\frac{1}{p_{11}}
$$
we have that the set of equations (\ref{yetanother1})-(\ref{yetanother2}) and (\ref{yetanother3})-(\ref{yetanother4}) are the same. Hence we conclude that Perron's classic eigenvalue problem is a particular case of the problem for $\mathcal{L}$ acting on matrices.

\qee

\section{A theorem on eigenvalues for the Ruelle operator}\label{sec_eigenvalues}

The following proposition is inspired in \cite{par}. We say that a hermitian operator  $P:V\to V$ on a Hilbert space $(V,\langle\cdot\rangle)$ is {\bf positive} if $\langle Pv,v\rangle\geq 0$, for all $v\in V$, denoted $P\geq 0$. Consider the positive operator $\mathcal{L}_{W,V}:\mathcal{PH}_N\to \mathcal{PH}_N$,
\begin{equation}
\mathcal{L}_{W,V}(\rho):=\sum_{i=1}^k tr(W_i\rho W_i^*)V_i\rho V_i^*
\end{equation}
We have the following result:
\begin{pro}\cite{BLLT}\label{proavl1}
There is $\rho\in\mathcal{M}_N$ and $\beta>0$ such that $\mathcal{L}_{W,V}(\rho)=\beta\rho$.
\end{pro}

\section{Vector integrals and barycenters}

We recall here a few basic definitions. For more details, see \cite{lozinski} and \cite{wsbook}. Let $X$ be a metric space. Let $(V,+,\cdot)$ be a real vector space, and $\tau$ a topology on $V$. We say that $(V,+,\cdot;\tau)$ is a topologic vector space if it is Hausdorff and if the operations $+$ and $\cdot$ are continuous. For instance, in the context of density matrices, we will consider $V$ as the Hilbert space $\mathcal{H}_N$ and $X$ will be the space of density matrices $\mathcal{M}_N$.

\begin{defi}
    Let $(X,\Sigma)$ be a measurable space, let $\mu\in M(X)$, let  $(V,+,\cdot ; \tau)$ be a locally convex space and let $f:X\to V$. we say that $x\in V$ is the {\bf integral} of $f$ in $X$, denoted by
$$x:=\int_X fd\mu$$
if
$$\Psi(x)=\int_X\Psi\circ fd\mu,$$
for all $\Psi\in V^*$.
\end{defi}

It is known that if we have a compact metric space $X$, $V$ is a
locally convex space and $f:X\to V$ is a continuous function such
that $\ov{co}f(X)$ is compact then the integral of $f$ in $X$
exists and belongs to $\ov{co}f(X)$. We will also use the
following well-known result, the barycentric formula:
\begin{pro}\label{winkler_teo}
\cite{winkler} Let $V$ be a locally convex space, let $E\subset V$ be a complete, convex and bounded set, and $\mu\in M^1(E)$.
Then there is a unique $x\in E$ such that
$$l(x)=\int_E l d\mu,$$
for all $l\in V^*$.
\end{pro}

\section{Example: density matrices}\label{sec_densities}

In this section we briefly review how the constructions of the previous section adjust to the case of density matrices.

\bigskip

Define $V:=\mathcal{H}_N$, $V^+:=\mathcal{PH}_N$ (note that such space is a convex cone), and let the partial order $\leq$ on $\mathcal{PH}_N$ be $\rho\leq\psi$ if and only if $\psi-\rho\geq 0$, i.e., if $\psi-\rho$ is positive. Then
$$(V,V^+,e)=(\mathcal{H}_N,\mathcal{PH}_N,tr),$$
is a regular state space \cite{wsbook}. Also, the set $B$ of unity trace in $V^+$ is, of course, the space of density matrices. Hence, $B=\mathcal{M}_N$.

\bigskip
Let $Z\subset V^*$ be a nonempty vector subspace of $V^*$. The smallest topology in $V$ such that every functional defined in $Z$ is continuous on that topology, denoted by $\sigma(V,Z)$, turns $V$ into a locally convex space. In particular, $\sigma(V,V^*)$ is the weak topology in $V$. If $(V,\Vert\cdot\Vert)$ is a normed space, then $\sigma(V^*,V)$ is called a weak$^*$ topology in $V^*$ (we identify $V$ with a subspace of $V^{**})$. We also have that $(C,\tau)=(\mathcal{PH}_N,\tau)$, where $\tau$ is the weak$^*$ topology (and which is equal to the Euclidean, see \cite{wsbook}) is a metrizable compact structure. In this case we have that $B_C=B\cap C=\mathcal{M}_N$.

\begin{defi}
A {\bf Markov operator} for probability measures is an operator $P:M^1(X)\to M^1(X)$ such that
$$P(\lambda\mu_1+(1-\lambda)\mu_2)=\lambda P\mu_1+(1-\lambda)P\mu_2,$$
for $\mu_1,\mu_2\in M^1(X)$, $\lambda\in (0,1)$.
\end{defi}
An example of such an operator is one which we have defined before and we denote it $\mathcal{V} : M^1(X)\to M^1(X)$,
\begin{equation}\label{markov_induced}
(\mathcal{V}\nu)(B)=\sum_{i=1}^k\int_{F_i^{-1}(B)}p_id\nu,
\end{equation}
and we call it the Markov operator induced by the IFS $\mathcal{F}$. We will be interested in fixed points for $\mathcal{V}$.

Define
$$m_b(X):=\{f:X\to\mathbb{R} : \textrm{f is bounded, measurable}\}$$
and also $\mathcal{U}:m_b(X)\to m_b(X)$,
$$(\mathcal{U}f)(x):=\sum_{i=1}^kp_i(x)f(F_i(x))$$

\begin{pro}\label{pdualbas}\cite{wsbook}
Let $f\in m_b(X)$ and $\mu\in M^1(X)$, then
$$\langle f,\mathcal{V}\mu\rangle=\langle\mathcal{U}f,\mu\rangle=\sum_{i=1}^k\int p_i(f\circ F_i)d\mu,$$
where $\langle f,\mu\rangle$ denotes the integral of $f$ with respect to $\mu$.
\end{pro}

\begin{defi} An operator $Q:V^+\to V^+$ is {\bf submarkovian} if
\begin{enumerate}
\item $Q(x+y)=Q(x)+Q(y)$
\item $Q(\alpha x)=\alpha Q(x)$
\item $\Vert Q(x)\Vert\leq\Vert x\Vert,$
\end{enumerate}
for all $x$, $y\in V^+$, $\alpha >0$.
\end{defi}
Every submarkovian operator $Q:V^+\to V^+$ can be extended in a unique way to a positive linear contraction on $V$.

\begin{defi} Let $P:V^+\to V^+$ a Markov operator and let
$P_i:V^+\to V^+$, $i=1,\dots, k$ be submarkovian operators such that $P=\sum_i P_i$. We say that $(P,\{P_i\}_{i=1}^k)$ is a {\bf Markov pair}.
\end{defi}
From \cite{wsbook}, we know that there is a 1-1 correspondence between homogeneous IFS and Markov pairs.

\begin{examp} In this example we want to obtain a probability $\eta$ such that $ \mathcal{V}(\eta)
=\eta$.

Suppose a QIFS, such that
$$p_i(\rho)=tr(W_i\rho W_i^*),\sp \sum_i W_i^* W_i=I,\sp F_i(\rho)=\frac{V_i\rho V_i^*}{tr(V_i\rho V_i^*)}$$
for $i=1,\dots, k$. Denote $m_b(\mathcal{M}_N)$ the space of bounded and measurable functions in $\mathcal{M}_N$. Consider  $\Lambda:\mathcal{M}_N\to \mathcal{M}_N$,
$$\Lambda(\rho)=\sum_i p_i(\rho)F_i(\rho)=\sum_i tr(W_i\rho W_i^*)\frac{V_i\rho V_i^*}{tr(V_i\rho V_i^*)}$$
Suppose there exists  a density matrix $\rho$ which $\Lambda$-invariant. As we know, such state is the barycenter of  $\mu$ which is  $\mathcal{V}$-invariant. Suppose $\mathcal{V}\mu=\mu$, then we can write
$$\int f d\mu=\int f d\mathcal{V}\mu=\sum_{i=1}^k \int p_i(\rho)f(F_i(\rho))d\mu(\rho)=\sum_i\int p_i(\rho)f\Big(\frac{V_i\rho V_i^*}{tr(V_i\rho V_i^*)}\Big)d\mu$$
$$=\sum_i\int tr(W_i\rho W_i^*)f\Big(\frac{V_i\rho V_i^*}{tr(V_i\rho V_i^*)}\Big)d\mu$$

Therefore, for any $f\in m_b(\mathcal{M}_N)$, we got the condition
\begin{equation}\label{n_eq_set09}
\int f d\mu=\sum_i\int tr(W_i\rho W_i^*)f\Big(\frac{V_i\rho V_i^*}{tr(V_i\rho V_i^*)}\Big)d\mu
\end{equation}

Let us consider a particular example where $N=2$, $k=4$, and
$$V_1=\left(
\begin{array}{cc}
\sqrt{p_{11}} & 0\\
0 & 0
\end{array}
\right),
\sp V_2=\left(
\begin{array}{cc}
0 & \sqrt{p_{12}}\\
0 & 0
\end{array}
\right)
,$$
$$V_3=\left(
\begin{array}{cc}
0 & 0\\
\sqrt{p_{21}} & 0
\end{array}
\right),
\sp V_4=\left(
\begin{array}{cc}
0 & 0\\
0 & \sqrt{p_{22}}
\end{array}
\right),
$$
 in such way that the $p_{ij}$ are the entries of a column stochastic  matrix $P$. Let $\pi=(\pi_1,\pi_2)$ be a vector such that  $P\pi=\pi$. A simple calculation shows that for $\rho$, the density matrix such that has entries $\rho_{ij}$, we have
\begin{equation}\label{300909a}
V_1\rho V_1^*=\left(
\begin{array}{cc}
p_{11}\rho_{11} & 0\\
0 & 0
\end{array}
\right),\sp V_2\rho V_2^*=\left(
\begin{array}{cc}
p_{12}\rho_{22} & 0\\
0 & 0
\end{array}
\right)
\end{equation}
\begin{equation}\label{300909b}
V_3\rho V_3^*=\left(
\begin{array}{cc}
0 & 0\\
0 & p_{21}\rho_{11}
\end{array}
\right),\sp V_4\rho V_4^*=\left(
\begin{array}{cc}
0 & 0\\
0 & p_{22}\rho_{22}
\end{array}
\right),
\end{equation}
and therefore
\begin{equation}
\frac{V_1\rho V_1^*}{tr(V_1\rho V_1^*)}=\left(
\begin{array}{cc}
1 & 0\\
0 & 0
\end{array}
\right) ,\sp \frac{V_2\rho V_2^*}{tr(V_2\rho V_2^*)}=\left(
\begin{array}{cc}
1 & 0\\
0 & 0
\end{array}
\right)
\end{equation}
\begin{equation}
\frac{V_3\rho V_3^*}{tr(V_3\rho V_3^*)}=\left(
\begin{array}{cc}
0 & 0\\
0 & 1
\end{array}
\right)  ,\sp \frac{V_4\rho V_4^*}{tr(V_4\rho V_4^*)}=\left(
\begin{array}{cc}
0 & 0\\
0 & 1
\end{array}
\right)
\end{equation}
that is, the above values do not depend on  $\rho$.

Define
\begin{equation}
\rho_x=\left(
\begin{array}{cc}
1 & 0\\
0 & 0
\end{array}
\right) ,\sp \rho_y=\left(
\begin{array}{cc}
0 & 0\\
0 & 1
\end{array}
\right)
\end{equation}
and
\begin{equation}
\eta=\pi_1\delta_{\rho_x}+\pi_2\delta_{\rho_y}
\end{equation}
Note that the barycenter of  $\eta$ is
$$\rho_{\eta}=\pi_1\rho_x+\pi_2\rho_y=\pi_1\left(
\begin{array}{cc}
1 & 0\\
0 & 0
\end{array}
\right)+\pi_2\left(
\begin{array}{cc}
0 & 0\\
0 & 1
\end{array}
\right)=\left(
\begin{array}{cc}
\pi_1 & 0\\
0 & \pi_2
\end{array}
\right)$$

For any  mensurable set $B$ we have
\begin{equation}
\mathcal{V}\eta(B)=\sum_{i=1}^4\int 1_B(F_i(\rho))p_i(\rho)d\eta=\sum_{i=1}^4\int 1_B\Big(\frac{V_i\rho V_i^*}{tr(V_i\rho V_i^*)}\Big)tr(V_i\rho V_i^*)d\eta
\end{equation}

We can now consider the following cases:
\begin{enumerate}
\item Suppose first that  $\rho_x$, $\rho_y\in B$. The using (\ref{300909a}) and (\ref{300909b}), one can show that
$$\mathcal{V}\eta(B)=\sum_{i=1}^4 \rho_{11}tr(V_i\rho_x V_i^*)+\rho_{22}tr(V_i\rho_y V_i^*)$$
$$=(\pi_1p_{11}+0)+(0+\pi_2p_{12})+(\pi_1p_{21}+0)+(0+\pi_2p_{22})=(\pi_1+\pi_2)=1,$$
because  $P\pi=\pi$.
\item Suppose now that $\rho_x\in B$, $\rho_y\notin B$
$$\mathcal{V}\eta(B)=\sum_{i=1}^4 \pi_1tr(V_i\rho_x V_i^*)=\pi_1(p_{11}+0+p_{21}+0)=\pi_1$$
\item Finally, suppose that $\rho_x\notin B$, $\rho_y\in B$
$$\mathcal{V}\eta(B)=\sum_{i=1}^4 \pi_2 tr(V_i\rho_y V_i^*)=\pi_2(0+p_{12}+0+p_{22})=\pi_2$$
\item It is easy to see that if $\rho_x, \rho_y\notin B$ then $\mathcal{V}\eta(B)=0$.
\end{enumerate}

The conclusion is that, $\mathcal{V}\eta(B)=\eta(B)$ for any measurable set  $B$.

Therefore,  $\mathcal{V}(\eta)=\eta$.

\qee
\end{examp}

\section{Some lemmas for IFS}\label{sec_somelemmas}

We want to understand the structure of $\Lambda:\mathcal{M}_N\to \mathcal{M}_N$,
$$\Lambda(\rho):=\sum_{i=1}^k p_iF_i=\sum_{i=1}^k tr(W_i\rho W_i^*)\frac{V_i\rho V_i^*}{tr(V_i\rho V_i^*)},$$
where $V_i$, $W_i$ are linear, $\sum_i W_i^*W_i=I$. Such operator is associated in a natural way to a IFS which is not homogeneous. In this section we state a few useful properties which are relevant for our study. The following lemmas hold for any IFS, except for lemma \ref{lema_imp01}, for which a proof is known for homogeneous IFS only.

\begin{lem}\label{umlemautil1}
Let $\{X,F_i,p_i\}_{i=1,\dots, k}$ be a IFS, $\Psi$ a linear functional on $X$. Then $\mathcal{U}\circ\Psi=\Psi\circ\Lambda$.
\end{lem}

\begin{cor}\label{ccc1}
Let $\mathcal{F}=(X,F_i,p_i)_{i=1,\dots, k}$ be a IFS and let $\rho_0\in X$. Then $\Lambda(\rho_0)=\rho_0$ if and only if $\mathcal{U}(\Psi(\rho_0))=\Psi(\rho_0)$, for all $\Psi$ linear functional.
\end{cor}

\begin{lem}\label{lemabas1}
Let $\mathcal{F}=\{X, F_i, p_i\}_{i=1,\dots, k}$ be a IFS.
\begin{enumerate}
\item Let $\rho_0\in X$ such that $F_i(\rho_0)=\rho_0$, $i=1,\dots, k$. Then $\mathcal{V}\delta_{\rho_0}=\delta_{\rho_0}$.
\item Let $\rho_0\in X$ such that $\mathcal{V}\delta_{\rho_0}=\delta_{\rho_0}$, then $\Lambda(\rho_0)=\rho_0$.
\end{enumerate}
\end{lem}

\begin{lem}\label{lema_imp01}
Let $\{X,F_i,p_i\}_{i=1,\dots, k}$ be a homogeneous IFS, $\Lambda=\sum_i p_i F_i$.

\begin{enumerate}
\item Let $\rho_{\nu}$ be the barycenter of a probability measure $\nu$. Then $\Lambda(\rho_{\nu})$ is the barycenter of $\mathcal{V}\nu$, where $\mathcal{V}$ is the associated Markov operator.

\item Let $\mu$ be an invariant probability measure for $\mathcal{V}$. Then the barycenter of $\mu$, denoted by $\rho_\mu$, is a fixed point of $\Lambda$.

\end{enumerate}

\end{lem}

\begin{examp}\label{exemplo_ifs} Let $k=N=2$,
$$V_1=\left(
\begin{array}{cc}
-1 & 0 \\
0 & 1
\end{array}
\right)
,\sp
V_2=\left(
\begin{array}{cc}
0 & -\frac{3\sqrt{2}}{4} \\
-\frac{3\sqrt{2}}{2} & 0
\end{array}
\right),$$
$W_1=(1/2)I$, $W_2=(\sqrt{3}/2)I$. Then
$$\Lambda(\rho)=\sum_i p_i(\rho)F_i(\rho)=\sum_i tr(W_i\rho W_i^*)\frac{V_i\rho V_i^*}{tr(V_i\rho V_i^*)}$$
$$=\frac{1}{4}V_1\rho V_1^*+\frac{3}{4}\frac{V_2\rho V_2^*}{tr(V_2\rho V_2^*)}
=\frac{1}{4}V_1\rho V_1^*+\frac{3}{4}\frac{V_2\rho V_2^*}{(\frac{9}{8}+\frac{27}{8}\rho_1)}$$
induces a IFS and it is such that  $\rho_0=\frac{1}{3}\vert 0\rangle\langle 0\vert+\frac{2}{3}\vert 1\rangle\langle 1\vert$ is a fixed point, with $F_1(\rho_0)=F_2(\rho_0)=\rho_0$. We can apply lemma \ref{lemabas1} and conclude that $\delta_{\rho_0}$ is an invariant measure for the Markov operator $\mathcal{V}$ associated to the IFS determined by $p_i$ and $F_i$.
\end{examp}

\qee

The following lemma, a simple variation from results seen in
\cite{wsbook}, determines reasonable conditions that we will need in
order to obtain a fixed point for $\mathcal{L}$ from a certain
measure which is invariant for the Markov operator $\mathcal{V}$.

\begin{lem}\label{ptofixapa}
Let $\{\mathcal{M}_N,F_i,p_i\}_{i=1,\dots, k}$ be an IFS which admits an attractive invariant measure $\mu$ for $\mathcal{V}$. Then $\lim_{n\to\infty}\Lambda^n(\rho_0)=\rho_{\mu}$, for every $\rho_0\in\mathcal{M}_N$, where $\rho_{\mu}$ is the barycenter of $\mu$.
\end{lem}

\section{Integral formulae for the entropy of IFS}\label{sec_entr_ifs}

Part of the results we present here in this section are variations
of the results presented in \cite{wsbook}. Let $(X,d)$ be a
complete separable metric space. Let $(V,V^+,e)$ be a complete
state space, $B=\{x\in V^+: e(x)=1\}$ and
$\mathcal{F}=(X,F_i,p_i)_{i=1,\dots, k}$ the homogeneous IFS
induced by the Markov pair $(\Lambda,\{\Lambda_i\}_{i=1}^k)$. Let
$I_{k}=\{1,\dots, k\}$ Let $n\in\mathbb{N}$, $\iota\in I_k^n$,
$i\in I_k$. Define $F_{\iota i}:=F_i\circ F_\iota$ and
\begin{equation}
p_{\iota i}(x)=\left\{\begin{array}{ll}
p_i(F_{\iota}x)p_{\iota}(x) & \textrm{ if } p_\iota(x)\neq 0\\
0 & \textrm{ otherwise }
\end{array}\right.
\end{equation}

\begin{pro}\label{un_etc}
Let $n\in\mathbb{N}$, $f\in m_b(X)$, $x\in X$. Then
$$(\mathcal{U}^n f)(x)=\sum_{\iota\in I_k^n} p_\iota(x)f(F_\iota(x))$$
\end{pro}

\begin{pro}
Let $x\in B$, $n\in\mathbb{N}$. Then
$$\Lambda^n(x)=\sum_{\iota\in I_k^n} p_\iota(x)F_\iota(x).$$
\end{pro}

\begin{pro}\label{ws_ident}
Let $\mathcal{F}$ be a IFS and let $g:B\to\mathbb{R}$. Then for $n\in\mathbb{N}$,
\begin{enumerate}
\item If g is concave (resp. convex, affine) then $\mathcal{U}^n g\leq g\circ \Lambda^n$ (resp. $\mathcal{U}^n g\geq g\circ \Lambda^n$, $\mathcal{U}^n g=g\circ\Lambda^n$).
\item If $\ov{x}$ is a fixed point for $\Lambda$ then the sequence $(\mathcal{U}^ng)(\ov{x}))_{n\in\mathbb{N}}$ is decreasing (resp. increasing, constant) if $g$ is concave (resp. convex, affine).

\bigskip
Also suppose that $\mathcal{F}$ is homogeneous. Then

\item If g is concave (resp. convex, affine), then $\mathcal{U}g$ is concave (resp. convex, affine).
\end{enumerate}
\end{pro}

\bigskip

Define $\eta:\mathbb{R}^+\to\mathbb{R}$ as

\begin{displaymath}
\eta(x)=\left\{\begin{array}{ll}
-x\log{x} & \textrm{ if } x\neq 0\\
0 & \textrm{ if } x=0
\end{array}\right.
\end{displaymath}

Define the {\bf Shannon-Boltzmann entropy function} as $h:X\to \mathbb{R}^+$,

$$h(x):=\sum_{i=1}^k\eta(p_i(x))$$
Let $n\in\mathbb{N}$. Define the {\bf partial entropy} $H_n:X\to\mathbb{R}^+$ as
$$H_n(x):=\sum_{\iota\in I_k^n}\eta(p_\iota (x)),$$
for $n\geq 1$ and $H_0(x):=0$, $x\in X$. Define, for $x\in X$,
$$\ov{\mathcal{H}}(x):=\limsup_{n\to\infty}\frac{1}{n}H_n(x),$$
the {\bf upper entropy on x}, and
$$\underline{\mathcal{H}}(x):=\liminf_{n\to\infty}\frac{1}{n}H_n(x),$$
the {\bf lower entropy on x}. If such limits are equal, we call its common value the {\bf entropy on x}, denoted by $\mathcal{H}(x)$.

\bigskip
Denote by $M^{\mathcal{V}}(X)$ the set of $\mathcal{V}$-invariant
probability measures on $X$. Let $\mu\in M^{\mathcal{V}}(X)$. The
{\bf partial entropy of the measure  $\mu$} is defined by
$$H_n(\mu):=\sum_{\iota\in I_k^n}\eta(\langle p_\iota,\mu\rangle),$$
for $n\geq 1$ and $H_0(\mu):=0$.

\begin{pro}
Let $\mu\in M^{\mathcal{V}}(X)$. Then the sequences
$(\frac{1}{n}H_n(\mu))_{n\in\mathbb{N}}$ and
$(H_{n+1}(\mu)-H_{n}(\mu))_{n\in\mathbb{N}}$ are nonnegative,
decreasing, and have the same limit.
\end{pro}
We denote the common limit of the sequences mentioned in the proposition above
as $\mathcal{H}(\mu)$ and we call it the {\bf entropy of the measure} $\mu$, i.e.,
$$\mathcal{H}(\mu):=\lim_{n\to\infty}\frac{1}{n}H_n(\mu)=\lim_{n\to\infty}(H_{n+1}(\mu)-H_{n}(\mu))$$

\bigskip

The following result gives us an integral formula for entropy, and
also a relation between the entropies defined before. We write
$S(\mu):=M^{\mathcal{V}}(X)\cap
\textrm{Lim}(\mathcal{V}^n\mu)_{n\in\mathbb{N}},$ where
$\textrm{Lim}(\mathcal{V}^n\mu)_{n\in\mathbb{N}}$ is the convex
hull of the set of accumulation points of
$(\mathcal{V}^n\mu)_{n\in\mathbb{N}}$, and $S_{\mathcal{F}}(\mu)$
is the set $S(\mu)$ associated to the Markov operator induced by
the IFS $\mathcal{F}$. For the definition of compact structure and
$(C,\tau)$-continuity, see \cite{wsbook}.

\begin{teo}\label{teo71entropia}
\cite{wsbook} (Integral formula for entropy of homogeneous IFS, compact case). Let $(C,\tau)$ be a metrizable compact structure $(V,V^+,e)$ such that $(\Lambda,\{\Lambda_i\}_{i=1}^k)$ is $(C,\tau)$-continuous. Assume that  $\rho_0\in B_C:=B\cap C$ is such that $\Lambda(\rho_0)=\rho_0$. Then
$$\mathcal{H}(\rho_0)=\mathcal{H}(\nu)=\int_X hd\nu$$
for each $\nu\in S_{\mathcal{F}_C}(\delta_{\rho_0})$, where $\mathcal{F}_C$ is the IFS $\mathcal{F}$ restricted to $(B_C,\tau)$.
\end{teo}

The analogous result for hyperbolic IFS is the following.

\begin{teo}\label{teo71entropiab}
\cite{wsbook} Let $\mathcal{F}=(X,F_i,p_i)_{i=1,\dots, k}$ be a hyperbolic IFS, $x\in X$, $\mu\in M^1(X)$ an invariant attractive measure for $\mathcal{F}$. Then
$$\mathcal{H}(x)=\lim_{n\to\infty}(H_{n+1}(x)-H_n(x))$$
and
$$\mathcal{H}(x)=\mathcal{H}(\mu)=\int_X hd\mu .$$
\end{teo}

\section{Some calculations on entropy}\label{sec_calc_entr}

Let $U$ be a unitary matrix of order $mn$ acting on $\mathcal{H}_m\otimes\mathcal{H}_n$. Its Schmidt decomposition is
$$U=\sum_{i=1}^K\sqrt{q_i}V_i^A\otimes V_i^B,\sp K=min\{m^2,n^2\}$$ The operators $V_i^A$ and $V_i^B$ act on certain Hilbert spaces $\mathcal{H}_m$ and $\mathcal{H}_n$, respectively. We also have that  $\sum_{i=1}^K q_i=1$. Let $\sigma=\rho_A\otimes\rho_*^B=\rho_A\otimes I_n/n$ and define
$$\Lambda(\rho_A):=tr_B(U\sigma U^*)=\sum_{i=1}^Kq_i V_i^A\rho_A V_i^{A*}$$
Recall that
$$tr_B(\vert a_1\rangle\langle a_2\vert\otimes \vert b_1\rangle\langle b_2\vert):
=\vert a_1\rangle\langle a_2\vert tr(\vert b_1\rangle\langle b_2\vert)$$
where $\vert a_1\rangle$ and $\vert a_2\rangle$ are vectors on the state space of $A$ and $\vert b_1\rangle$ and $\vert b_2\rangle$ are vectors on the state space of $B$. The trace on the right side is the usual trace on $B$. A calculation shows that if $\rho_*^A=I_m/m$, then $\Lambda(\rho_*^A)=\rho_*^A$ and so $\Lambda$ is such that $\Lambda(I_m/m)=I_m/m$ and $\Lambda$ is trace preserving.

\bigskip

Let $\mathcal{F}$ be the homogeneous IFS associated to the $V_i^A$, that is,  $p_i(\rho)=tr(q_iV_i^A\rho V_i^{A*})$, $F_i(\rho)=(q_iV_i^A\rho V_i^{A*})/tr(q_iV_i^A\rho V_i^{A*})$ and let $\rho_0$ be a fixed point of $\Lambda=\sum_i p_iF_i$. Following \cite{wsbook}, we have that $\rho_0$ is the barycenter of $\mathcal{V}^n\delta_{\rho_0}$, $n\in\mathbb{N}$. By theorem \ref{teo71entropia}, we can calculate  the entropy of such IFS. In this case we have
\begin{equation}\label{expressao_entr}
\mathcal{H}(\rho_0)=\mathcal{H}(\nu)=\int_{\mathcal{M}_N} h d\nu,
\end{equation}
where $\nu\in M^{\mathcal{V}}(X)\cap
\textrm{Lim}(\mathcal{V}^n\delta_{\rho_0})_{n\in\mathbb{N}}$.

\qee

Let $\mathcal{F}=(\mathcal{M}_N,F_i,p_i)_{i=1,\dots, k}$ be an IFS, $\Lambda(\rho)=\sum_i p_iF_i$. Let $\mathcal{U}$ be the conjugate of $\mathcal{V}$. By proposition \ref{un_etc},
$$(\mathcal{U}^n h)(\rho)=\sum_{\iota\in I_k^n(\rho)} p_\iota(\rho)h(F_\iota(\rho))$$
and since $h(\rho)=\sum_{j=1}^k\eta(p_j(\rho))$, we have, for $\iota=(i_1,\dots, i_n)$, and every $\rho_0\in\mathcal{M}_N$,
\begin{equation}\label{inicio_copia}
\int_{\mathcal{M}_N} h d\mathcal{V}^n\delta_{\rho_0}=\int_{\mathcal{M}_N} \mathcal{U}^n h d\delta_{\rho_0}
\end{equation}
\begin{equation}
=-\int_{\mathcal{M}_N}\sum_{\iota\in I_k^n(\rho)}p_\iota(\rho)\sum_{j=1}^k p_j(F_\iota(\rho))\log{p_j(F_\iota(\rho))}d\delta_{\rho_0}
\end{equation}
\begin{equation}
=-\sum_{\iota\in I_k^n(\rho_0)}p_\iota(\rho_0)\sum_{j=1}^k p_j(F_\iota(\rho_0))\log{p_j(F_\iota(\rho_0))}
\end{equation}
\begin{equation}
=-\sum_{\iota\in I_k^n(\rho_0)}p_{i_1}(\rho_0)p_{i_2}(F_{i_1}\rho_0)\cdots p_{i_n}(F_{i_{n-1}}(F_{i_{n-2}}(\cdots (F_{i_1}\rho_0))))\times
\end{equation}
\begin{equation}\label{ac_conta1}
\times\sum_{j=1}^k p_j(F_{i_n}(F_{i_{n-1}}(\cdots(F_{i_1}\rho_0))))\log{p_j(F_{i_n}(F_{i_{n-1}}(\cdots(F_{i_1}\rho_0))))}=(\mathcal{U}^n h)(\rho_0)
\end{equation}

Suppose $\Lambda(\rho_0)=\rho_0$. We have by proposition \ref{ws_ident}, since $h$ is concave, that $(\mathcal{U}^nh)_{n\in\mathbb{N}}$ is decreasing, $\mathcal{U}^nh\leq h\circ\Lambda^n$ and so
\begin{equation}\label{ac_conta2}
\int_{\mathcal{M}_N} h d\mathcal{V}^n\delta_{\rho_0}\leq h(\Lambda^n(\rho_0))= h(\rho_0),
\end{equation}
for every $n$.

\section{An expression for a stationary entropy}\label{sec_novaentr}

In this section we present a definition of entropy which captures a stationary behavior.

Let $H$ be a hermitian operator and $V_i$, $i=1,\dots, k$ linear operators. We can define the dynamics $F_i:\mathcal{M}_N\to \mathcal{M}_N$:
\begin{equation}\label{din01}
F_i(\rho):=\frac{V_i\rho V_i^*}{tr(V_i\rho V_i^*)}
\end{equation}
Let $W_i$, $i=1,\dots, k$ be linear and such that $\sum_{i=1}^k W_i^*W_i=I$. This determines functions $p_i:\mathcal{M}_N\to\mathbb{R}$,
\begin{equation}\label{prob01}
p_i(\rho):=tr(W_i\rho W_i^*)
\end{equation}
Then we have $\sum_{i=1}^k p_i(\rho)=1$, for every $\rho$. Therefore a family $W:=\{W_i\}_{i=1,\dots, k}$ determines a QIFS $\mathcal{F}_{W}$,
$$\mathcal{F}_W=\{\mathcal{M}_N,F_i,p_i\}_{i=1,\dots, k}$$
with $F_i$, $p_i$ given by (\ref{din01}) and (\ref{prob01}).

 Different choices of $W_i, i=1,2...,k$, as above, determine different invariant probabilities.

 We introduce the following definition of entropy

\bigskip

\begin{defi} Suppose that we have a QIFS such that there is a unique attractive invariant measure for the Markov operator $\mathcal{V}$ associated to $\mathcal{F}_W$. Let $\rho_W$ be the barycenter of such measure. Define

\begin{equation}\label{nossa_entropia}
h_V(W):=-\sum_{i=1}^k tr(W_i\rho_W W_i^*)\sum_{j=1}^k tr\Big(\frac{W_j V_i\rho_W V_i^* W_j^*}{tr(V_i\rho_W V_i^*)}\Big)\log{tr\Big(\frac{W_j V_i\rho_W V_i^* W_j^*}{tr(V_i\rho_W V_i^*)}\Big)}
\end{equation}

\end{defi}

\bigskip

Remember that by lemma \ref{ptofixapa}, we have that $\rho_W$ is a fixed point for  \begin{equation}\label{lhat1}
\widehat{\mathcal{L}}_{\mathcal{F}_W}(\rho):=\sum_{i=1}^k p_i(\rho)F_i(\rho)=\sum_{i=1}^k tr(W_i\rho W_i^*)\frac{V_i\rho V_i^*}{tr(V_i\rho V_i^*)}
\end{equation}

\begin{lem} We have that $0\leq h_V(W)\leq \log k$, for every family $W_i$ of linear operators satisfying $\sum_{i=1}^k W_i^* W_i=I$. Also, for any given dynamics $V$ the maximum can be reached.
\end{lem}

We also define
\begin{equation}\label{lnohat1}
\mathcal{L}_{\mathcal{F}_W}(\rho):=\sum_{i=1}^k tr(W_i\rho W_i^*)V_i\rho V_i^*
\end{equation}

Note that by the construction made on section \ref{sec_novaentr}, we have $h_V(W)=\mathcal{U}h(\rho_W)$, where $\mathcal{U}h(\rho)=\sum_i p_i(\rho)h(F_i(\rho))$.

\qee

\begin{lem}
Let $\mathcal{F}=(\mathcal{M}_N,F_i,p_i)$ be a QIFS, with $F_i$, $p_i$ in the form (\ref{din01}) and (\ref{prob01}). Suppose there is $\rho_0\in\mathcal{M}_N$ such that $\delta_{\rho_0}$ is the unique $\mathcal{V}$-invariant measure. Then $\widehat{\mathcal{L}}_{\mathcal{F}}(\rho_0)=\rho_0$ (eq. (\ref{lhat1})) and
$$\int \mathcal{U}^n h d\delta_{\rho_0}=\mathcal{U}^nh(\rho_0)=h(\rho_0),$$
for all $n\in\mathbb{N}$. Besides, $\mathcal{U}^nh(\rho_0)=\mathcal{U}h(\rho_0)$ and so
$$h_V(W)=\mathcal{U}^n h(\rho_0),$$
for all $n\in\mathbb{N}$.
\end{lem}

\begin{lem}
Let $\mu$ be a $\mathcal{V}$-invariant attractive measure. Then if $\rho_{\mu}$ is the barycenter of $\mu$ we have, for any $\rho$,
\begin{equation}\label{wcheck}
\lim_{n\to\infty}\mathcal{U}^nh(\rho)=\int \mathcal{U}h d\mu=\int h d\mu\leq h(\rho_{\mu})
\end{equation}
\end{lem}

\begin{lem}
Let $\mathcal{F}=(\mathcal{M}_N,F_i,p_i)$ be a QIFS, with $F_i$, $p_i$ in the form (\ref{din01}) and (\ref{prob01}).
Suppose that $\rho$ is the unique point such that $\widehat{\mathcal{L}}_{\mathcal{F}}(\rho)=\rho$. Suppose that $F_i(\rho)=\rho$, $i=1,\dots, k$. Then
$$\mathcal{U}^nh(\rho)=h(\rho),$$
$n=1,2,\dots$, and therefore $h_V(W)$ does not depend on $n$.
\end{lem}

\section{Entropy and Markov chains}\label{camarkov}

Let $V_i$, $W_i$ be linear operators, $i=1,\dots, k$, $\sum_{i=1}^k W_i^*W_i=I$. Suppose the $V_i$ are fixed and determine a dynamics given by $F_i:\mathcal{M}_N\to \mathcal{M}_N$, $i=1,\dots, k$. Define
$$P:=\{(p_1,\dots, p_k): p_i:\mathcal{M}_N\to\mathbb{R}^+,i=1,\dots, k ,\sum_{i=1}^k p_i(\rho)=1,\forall\rho\in\mathcal{M}_N\}$$
$$P':=P\cap\{(p_1,\dots, p_k):\exists W_i, i=1,\dots, k: p_i(\rho)=tr(W_i\rho W_i^*),$$
$$ W_i \textrm{ linear }, \sum_i W_i^*W_i=I\}$$

$$\mathcal{M}_F:=\{\mu\in M^1(\mathcal{M}_N): \exists p\in P' \textrm{ such that } \mathcal{V}_p\mu=\mu\},$$
where $\mathcal{V}_p:M^1(\mathcal{M}_N)\to M^1(\mathcal{M}_N)$,

$$\mathcal{V}_p(\mu)(B):=\sum_{i=1}^k\int_{F_i^{-1}(B)}p_id\mu$$

Note that a family $W:=\{W_i\}_{i=1,\dots, k}$ determines a QIFS $\mathcal{F}_{W}$,
$$\mathcal{F}_W=\{\mathcal{M}_N,F_i,p_i\}_{i=1,\dots, k}$$

As done in the previous section we introduce the following definition (which is in some sense stationary)
\begin{equation}
h_V(W):=-\sum_{i=1}^k \frac{tr(W_i\rho_W W_i^*)}{tr(V_i\rho_W V_i^*)}\sum_{j=1}^k tr\Big(W_j V_i\rho_W V_i^*
W_j^*\Big)\log{\Big(\frac{tr(W_j V_i\rho_W V_i^* W_j^*)}{tr(V_i\rho_W V_i^*)}\Big)}
\end{equation}
\normalsize{where} as before, $\rho_W$ denotes the barycenter of the unique attractive invariant measure for the Markov operator $\mathcal{V}$ associated to $\mathcal{F}_W$.

Let $P=(p_{ij})_{i,j=1,\dots,N}$ be a stochastic, irreducible matrix. Let $p$ be the stationary vector of $P$. The entropy of $P$ is defined as
\begin{equation}\label{eestoc}
H(P):=-\sum_{i,j=1}^Np_i p_{ij}\log{p_{ij}}
\end{equation}
We consider an example which shows that the usual Markov chain entropy can be realized as the entropy associated to a certain QIFS.

\begin{examp} (Homogeneous case, 4 matrices).
Let $N=2$, $k=4$ and
$$V_1=\left(
\begin{array}{cc}
\sqrt{p_{00}} & 0\\
0 & 0
\end{array}
\right),
\sp V_2=\left(
\begin{array}{cc}
0 & \sqrt{p_{01}}\\
0 & 0
\end{array}
\right)
,$$
$$V_3=\left(
\begin{array}{cc}
0 & 0\\
\sqrt{p_{10}} & 0
\end{array}
\right),
\sp V_4=\left(
\begin{array}{cc}
0 & 0\\
0 & \sqrt{p_{11}}
\end{array}
\right)
$$
Note that
$$\sum_i V_i^* V_i=\left(
\begin{array}{cc}
p_{00}+p_{10} & 0\\
0 & p_{01}+p_{11}
\end{array}
\right)$$
and so $\sum_i V_i^* V_i=I$ if we suppose that
$$P:=\left(
\begin{array}{cc}
p_{00} & p_{01}\\
p_{10} & p_{11}
\end{array}
\right)$$
is column-stochastic. We have
$$V_1\rho V_1^*=\left(
\begin{array}{cc}
p_{00}\rho_1 & 0\\
0 & 0
\end{array}
\right),\sp
V_2\rho V_2^*=\left(
\begin{array}{cc}
p_{01}\rho_4 & 0\\
0 & 0
\end{array}
\right)$$
$$V_3\rho V_3^*=\left(
\begin{array}{cc}
0 & 0\\
0 & p_{10}\rho_1
\end{array}
\right),\sp
V_4\rho V_4^*=\left(
\begin{array}{cc}
0 & 0\\
0 & p_{11}\rho_4
\end{array}
\right)
$$
so
$$tr(V_1\rho V_1^*)=p_{00}\rho_1,\sp tr(V_2\rho V_2^*)=p_{01}\rho_4$$
$$\sp tr(V_3\rho V_3^*)=p_{10}\rho_1,\sp tr(V_4\rho V_4^*)=p_{11}\rho_4$$

\bigskip
The fixed point of $\Lambda(\rho)=\sum_i V_i\rho V_i^*$ is
$$\rho_V=\left(
\begin{array}{cc}
\frac{p_{01}}{1-p_{00}+p_{01}} & 0\\
0 & \frac{1-p_{00}}{1-p_{00}+p_{01}}
\end{array}
\right)
$$

Let $\pi=(\pi_1,\pi_2)$ such that $P\pi=\pi$. We know that
\begin{equation}
\pi=(\frac{p_{01}}{1-p_{00}+p_{01}},\frac{1-p_{00}}{1-p_{00}+p_{01}})
\end{equation}
Then the nonzero entries of $\rho_V$ are the entries of $\pi$ and
so we associate the fixed point of $P$ to the fixed point of a
certain $\Lambda$ in a natural way. Let us calculate  $h_V(W)$.
Note that $\Lambda$ defined above is associated to a homogeneous
IFS. Then $W_i=V_i$, $i=1,\dots, k$ and
$$h_V(W)=h_V(V)$$
$$=-\sum_{i=1}^k \frac{tr(W_i\rho_V W_i^*)}{tr(V_i\rho_V V_i^*)}\sum_{j=1}^k tr\Big(W_j V_i\rho_V V_i^* W_j^*\Big)\log{\Big(\frac{tr(W_j V_i\rho_V V_i^* W_j^*)}{tr(V_i\rho_V V_i^*)}\Big)}$$
\begin{equation}
=-\sum_{i,j}tr\Big(V_j V_i\rho_V V_i^* V_j^*\Big)\log{\Big(\frac{tr(V_j V_i\rho_V  V_i^* V_j^*)}{tr(V_i\rho_V V_i^*)}\Big)}
\end{equation}
A simple calculation yields $H(P)=h_V(V)$, where $H(P)$ is the entropy of $P$, given by (\ref{eestoc}). This shows that the entropy of Markov chains is a particular case of the entropy for QIFS defined before.
\end{examp}

\qee

In a similar way, we can reach the same conclusion for the nonhomogeneous case, 4 matrices, and also for 2 matrices \cite{BLLT}.

\qee

\begin{lem}
Let $V_{ij}$ be matrices of order $n$,
$$V_{ij}=\sqrt{p_{ij}}\vert i\rangle\langle j\vert$$
for $i,j=1,\dots, n$. Let $$\Lambda_P(\rho):=\sum_{i,j}V_{ij}\rho V_{ij}^*$$
where $P=(p_{ij})_{i,j=1,\dots, n}$. Then for all $n$, $\Lambda_P^n(\rho)=\Lambda_{P^n}(\rho)$.
\end{lem}

\begin{cor}
Under the lemma hypothesis, we have $\lim_{n\to\infty}\Lambda_P^n(\rho)=\Lambda_{\pi}(\rho)$, where $\pi=\lim_{n\to\infty}P^n$ is the stochastic matrix which has all columns equal to the stationary vector for $P$.
\end{cor}

\section{Capacity-cost function and pressure}\label{sec_cc}

Recall that every trace preserving, completely positive (CP) mapping can be written in the Stinespring-Kraus form,
$$\Lambda(\rho)=\sum_{i=1}^k V_i\rho V_i^*,\sp \sum_{i=1}^k V_i^*V_i=I,$$
for $V_i$ linear operators. These mappings are also called {\bf quantum channels}.

This is one of the main motivations for considering the class of
operators (a generalization of the above ones) described in the
present work. These are natural objets in the study of Quantum
Computing.

\begin{defi} The {\bf Holevo capacity} for sending classic information via a quantum channel $\Lambda$ is defined as
\begin{equation}
C_{\Lambda}:=\max_{\stackrel{p_i\in [0,1]}{\rho_i\in\mathcal{M}_N}}S\Big(\sum_{i=1}^n p_i\Lambda(\rho_i)\Big)-\sum_{i=1}^n p_i S\Big(\Lambda(\rho_i)\Big)
\end{equation}
where $S(\rho)=-tr(\rho\log\rho)$ is the von Neumann entropy. The
maximum is, therefore, over all choices of $p_i$, $i=1,\dots, n$
and density operators $\rho_i$, for some $n\in \mathbb{N}$. The
Holevo capacity establishes an upper bound on the amount of
information that a quantum system contains \cite{nich}.
\end{defi}

\begin{defi} Let $\Lambda$ be a quantum channel. Define the {\bf minimum output entropy} as
$$H^{min}(\Lambda):=\min_{\vert\psi\rangle}S(\Lambda(\vert\psi\rangle\langle\psi\vert))$$
\end{defi}

{\bf Additivity conjecture} We have that
$$C_{\Lambda_1\otimes\Lambda_2}=C_{\Lambda_1}+C_{\Lambda_2}$$

{\bf Minimum output entropy conjecture} For any channels $\Lambda_1$ and $\Lambda_2$,
$$H^{min}(\Lambda_1\otimes\Lambda_2)=H^{min}(\Lambda_1)+H^{min}(\Lambda_2)$$

\bigskip
In \cite{shor}, is it shown that the additivity conjecture is equivalent to the minimum output entropy conjecture, and in \cite{hastings} we obtain a counterexample for this last conjecture.

\qee

We will be interested here in a different class of problem which
concern maximization (and not minimization) of entropy plus  a
given potential (a cost) \cite{gray}, \cite{hayashi1},
\cite{hayashi2}.

\begin{defi}
Let $M_F$ be the set of invariant measures defined in the section \ref{camarkov} and let $H$ be a hermitian operator. For $\mu\in\mathcal{M}_F$ let $\rho_{\mu}$ be its barycenter. Define the capacity-cost function $C:\mathbb{R}^+\to\mathbb{R}^+$ as
\begin{equation}
C(a):=\max_{\mu\in\mathcal{M}_F} \{ h_{W,V}(\rho_{\mu}): tr(H\rho_{\mu})\leq a\}
\end{equation}
\end{defi}
The following analysis is inspired in \cite{lopescra}. There is a relation between the cost-capacity function and the variational problem for pressure. In fact, let $F:\mathbb{R}^+\to\mathbb{R}^+$ be the function given by
\begin{equation}
F(\lambda):=\sup_{\mu\in\mathcal{M}_F}\{h_{W,V}(\rho_{\mu})-\lambda tr(H\rho_{\mu})\}
\end{equation}
We have the following fact. There is a unique probability measure $\nu_0\in\mathcal{M}_F$ such that
$$F(\lambda)=h_{W,V}(\rho_{\nu_0})-\lambda tr(H\rho_{\nu_0})$$
Also, we have the following lemma:
\begin{lem}
Let $\lambda\leq 0$, and $\hat{a}=tr(H\rho_{\nu_0})$. Then
\begin{equation}
C(\hat{a})=h_{W,V}(\rho_{\nu_0})
\end{equation}
\end{lem}

\section{Analysis of the pressure problem}\label{sec_analysis1}

Let $V_i$, $W_i$ be linear operators, $i=1,\dots, k$, with $\sum_i W_i^*W_i=I$ and let
\begin{equation}\label{um_pot_int}
H\rho:=\sum_{i=1}^k H_i\rho H_i^*
\end{equation}
a hermitian operator. We are interested in obtaining a version of the variational principle of pressure for our context. We will see that the pressure will be maximum whenever we have a certain relation between the potential $H$ and the probability distribution considered (and represented here by the $W_i$). Initially we consider that the $V_i$ are fixed. From the reasoning described below, it will be natural to consider as definition of pressure the maximization among the possible stationary $W_i$ of the expression
$$
h_V(W)+ \sum_{j=1}^k \log \Big( tr(H_j\rho_{\beta}  H_j^*)tr(V_j\rho_{\beta} V_j^*)\Big) tr(W_j \rho_W  W_j^*)$$

Remember that different choices of $W_i, i=1,2,...,k$, represent different choices of invariant probabilities.

Our analysis uses the following important lemma.

\begin{lem}\label{lemalog2}
 If $r_1,\dots , r_k$ and $q_1,\dots ,q_k$ are two probability distributions over $1,\dots, k$,   such that $r_j>0$, $j=1,\dots, k$, then
$$-\sum_{j=1}^kq_j\log{q_j}+\sum_{j=1}^k q_j\log{r_j}\leq 0$$
and equality holds if and only if $r_j=q_j$, $j=1,\dots, k$.
\end{lem}
For the proof, see \cite{par}.

\bigskip

The potential given by (\ref{um_pot_int}) together with the $V_i$ induces an operator, given by
\begin{equation}\label{operador_usa1}
\mathcal{L}_H(\rho):=\sum_{i=1}^k tr(H_i\rho H_i^*)V_i\rho V_i^*
\end{equation}
We know that such operator admits an eigenvalue $\beta$ with its associate eigenstate $\rho_{\beta} $. Then $\mathcal{L}_H(\rho_{\beta} )=\beta\rho_{\beta} $ implies
\begin{equation}\label{soma_1_a}
\sum_{i=1}^k tr(H_i\rho_{\beta}  H_i^*)V_i\rho_{\beta}  V_i^*=\beta\rho_{\beta}
\end{equation}
In coordinates, (\ref{soma_1_a}) can be written as
\begin{equation}\label{soma_1_b}
\sum_{i=1}^k tr(H_i\rho_\beta H_i^*)(V_i\rho_\beta V_i^*)_{jl}=\beta(\rho_\beta)_{jl}
\end{equation}

\bigskip

{\bf Remark} Comparing the above calculation with the problem of
finding an eigenvalue $\lambda$ of a matrix $A=(a_{ij})$, we have
that equation (\ref{soma_1_a}) can be seen as the analogous of the
expression
\begin{equation}
lE^A=\lambda l
\end{equation}
Above, the matrix $A$ plays the role of a potential, $E^A$ denotes
the matrix with entries $e^{a_{ij}}$ and $l_j$ denotes the $j$-th
coordinate of the left eigenvector $l$ associated to the
eigenvalue $\lambda$. In coordinates,
\begin{equation}
\sum_i l_i e^{a_{ij}}=\lambda l_j, \sp i,j=1,\dots, k
\end{equation}
\qee

From this point we can perform two calculations. First,
considering (\ref{soma_1_a}) we will take the trace of such
equation in order to obtain a scalar equation. In spite of the
fact that taking the trace makes us lose part of the information
given by the eigenvector equation, we are still able to obtain a
version of what we will call a {\bf basic inequality}, which can
be seen as a quantum IFS version of the variational principle of
pressure. However, there is an algebraic drawback to this
approach, namely, that we will not be able to have the classic
variational problem as a particular case of such inequality (such
disadvantage is a consequence of taking the trace, clearly). The
second calculation will consider (\ref{soma_1_b}), the coordinate
equations associated to the matrix equation for the eigenvectors.
In this case we also obtain a basic inequality, but now we will
have the classic variational problem of pressure as a particular
case.

\bigskip

An important question which is of our interest, regarding both calculations mentioned above, is the question of whether it is possible for a given system to attain its maximum pressure. It is not clear that given any dynamics, we can obtain a measure reaching such a maximum. With respect to our context, we will state sufficient conditions on the dynamics which allows us to determine expressions for the measure which maximizes the pressure. We now perform the calculations mentioned above.

\bigskip
Based on (\ref{soma_1_a}), define
\begin{equation}
r_j=\frac{1}{\beta}tr(H_j\rho_{\beta}  H_j^*)tr(V_j\rho_{\beta}  V_j^*)
\end{equation}
So we have $\sum_j r_j=1$. Let
\begin{equation}
q_j^i:=tr\Big(\frac{W_j V_i\rho_W V_i^* W_j^{*}}{tr(V_i\rho_W V_i^*)}\Big)
\end{equation}
where, as before, $\rho_W$ is the fixed point associated to the
renormalized operator $\Lambda_{\mathcal{F}_W}$,
\begin{equation}
\Lambda_{\mathcal{F}_W}(\rho):=\sum_{i=1}^k p_i(\rho)F_i(\rho)
\end{equation}
induced by the QIFS $(\mathcal{M}_N,F_i,p_i)_{i=1,\dots, k}$,
$$F_i(\rho)=\frac{V_i\rho V_i^*}{tr(V_i\rho V_i^*)}$$
and
$$p_i(\rho)=tr(W_i\rho W_i^*)$$
Note that we have
$$\sum_{j=1}^kq_j^i=\frac{1}{tr(V_i\rho_W V_i^*)}\sum_{j=1}^k tr(W_j^{*}W_j V_i\rho_W V_i^*)$$
$$=\frac{1}{tr(V_i\rho_W V_i^*)}tr(\sum_{j=1}^k W_j^{*}W_j V_i\rho_W V_i^*) =1$$

Then we can apply lemma \ref{lemalog2} for $r_j$, $q_j^i$, $j=1,\dots k$, with $i$ fixed, to obtain
$$-\sum_j tr\Big(\frac{W_j V_i\rho_W V_i^* W_j^{*}}{tr(V_i\rho_W V_i^*)}\Big)\log tr\Big(\frac{W_j V_i\rho_W V_i^* W_j^{*}}{tr(V_i\rho_W V_i^*)}\Big)$$
\begin{equation}
+\sum_j tr\Big(\frac{W_j V_i\rho_W V_i^* W_j^{*}}{tr(V_i\rho_W V_i^*)}\Big)\log \Big(\frac{1}{\beta}tr(H_j\rho_{\beta}  H_j^*)tr(V_j\rho_{\beta}  V_j^*)\Big)\leq 0
\end{equation}
and equality holds if and only if for all $i,j$,
\begin{equation}
\frac{1}{\beta}tr(H_j\rho_{\beta}  H_j^*)tr(V_j\rho_{\beta}  V_j^*)=\frac{tr(W_j V_i\rho_W V_i^* W_j^{*})}{tr(V_i\rho_W V_i^*)}
\end{equation}
Then
$$-\sum_j tr\Big(\frac{W_j V_i\rho_W V_i^* W_j^{*}}{tr(V_i\rho_W V_i^*)}\Big)\log tr\Big(\frac{W_j V_i\rho_W V_i^* W_j^{*}}{tr(V_i\rho_W V_i^*)}\Big)$$
$$+\sum_j tr\Big(\frac{W_j V_i\rho_W V_i^* W_j^{*}}{tr(V_i\rho_W V_i^*)}\Big)\log \Big( tr(H_j\rho_{\beta}  H_j^*)tr(V_j\rho_{\beta}  V_j^*)\Big)$$
$$\leq \sum_j tr\Big(\frac{W_j V_i\rho_W V_i^* W_j^{*}}{tr(V_i\rho_W V_i^*)}\Big)\log\beta
$$
which is equivalent to
$$-\sum_j tr\Big(\frac{W_j V_i\rho_W V_i^* W_j^{*}}{tr(V_i\rho_W V_i^*)}\Big)\log tr\Big(\frac{W_j V_i\rho_W V_i^* W_j^{*}}{tr(V_i\rho_W V_i^*)}\Big)$$
\begin{equation}
+\sum_j \frac{tr(W_j V_i\rho_W V_i^* W_j^{*})}{tr(V_i\rho_W V_i^*)}\log \Big( tr(H_j\rho_{\beta}  H_j^*)tr(V_j\rho_{\beta}  V_j^*)\Big)\leq \log\beta
\end{equation}
Multiplying by $tr(W_i\rho_W W_i^*)$ and summing over the $i$ index, we have
$$h_V(W)+\sum_j \log \Big( tr(H_j\rho_{\beta}  H_j^*)tr(V_j\rho_{\beta}  V_j^*)\Big) \sum_i \frac{tr(W_i\rho_W W_i^*)}{tr(V_i\rho_W V_i^*)} tr(W_j V_i\rho_W V_i^* W_j^{*})$$
\begin{equation}\label{pv_pr01xx}
\leq \sum_i tr(W_i\rho_W W_i^*)\log\beta=\log\beta
\end{equation}
and equality holds if and only if for all $i,j$,
\begin{equation}
\frac{1}{\beta}tr(H_j\rho_{\beta}  H_j^*)tr(V_j\rho_{\beta}  V_j^*)=\frac{tr(W_j V_i\rho_W V_i^* W_j^{*})}{tr(V_i\rho_W V_i^*)}
\end{equation}

Let us rewrite inequality (\ref{pv_pr01xx}). First we use the fact
that $\rho_W$ is a fixed point of $\Lambda_{\mathcal{F}_W}$,
\begin{equation}
\sum_{i=1}^k tr(W_i\rho_W W_i^*)\frac{V_i\rho_W V_i^*}{tr(V_i\rho_W V_i^*)}=\rho_W
\end{equation}
Now we compose both sides of the equality above with the operator
\begin{equation}
\sum_{j=1}^k \log \Big( tr(H_j\rho_{\beta}  H_j^*)tr(V_j\rho_{\beta}  V_j^*)\Big) W_j^*W_j
\end{equation}
and then we obtain
$$\sum_{i=1}^k tr(W_i\rho_W W_i^*)\frac{V_i\rho_W V_i^*}{tr(V_i\rho_W V_i^*)}\sum_{j=1}^k \log \Big( tr(H_j\rho_{\beta}  H_j^*)tr(V_j\rho_{\beta}  V_j^*)\Big) W_j^*W_j$$
\begin{equation}
=\rho_W \sum_{j=1}^k \log \Big( tr(H_j\rho_{\beta}  H_j^*)tr(V_j\rho_{\beta}  V_j^*)\Big) W_j^*W_j
\end{equation}
Reordering terms we get
$$\sum_{j=1}^k \log \Big( tr(H_j\rho_{\beta}  H_j^*)tr(V_j\rho_{\beta}  V_j^*)\Big) \sum_{i=1}^k \frac{tr(W_i\rho_W W_i^*)}{tr(V_i\rho_W V_i^*)} V_i\rho_W V_i^* W_j^*W_j$$
\begin{equation}
=\rho_W \sum_{j=1}^k \log \Big( tr(H_j\rho_{\beta}  H_j^*)tr(V_j\rho_{\beta}  V_j^*)\Big) W_j^*W_j
\end{equation}
Taking the trace on both sides we get
$$\sum_{j=1}^k \log \Big( tr(H_j\rho_{\beta}  H_j^*)tr(V_j\rho_{\beta}  V_j^*)\Big) \sum_{i=1}^k \frac{tr(W_i\rho_W W_i^*)}{tr(V_i\rho_W V_i^*)} tr(W_j V_i\rho_W V_i^* W_j^*)$$
\begin{equation}\label{subst_aquixx}
=\sum_{j=1}^k \log \Big( tr(H_j\rho_{\beta}  H_j^*)tr(V_j\rho_{\beta} V_j^*)\Big) tr(\rho_W  W_j^*W_j)
\end{equation}
Note that the left hand side of (\ref{subst_aquixx}) is one of the sums appearing in (\ref{pv_pr01xx}). Therefore replacing (\ref{subst_aquixx}) into (\ref{pv_pr01xx}) gives us the following inequality:
\begin{equation}\label{pvar_ver1xx}
h_V(W)+ \sum_{j=1}^k \log \Big( tr(H_j\rho_{\beta}  H_j^*)tr(V_j\rho_{\beta} V_j^*)\Big) tr(W_j \rho_W  W_j^*)
 \leq \log\beta
\end{equation}
and equality holds if and only if for all $i,j$,
\begin{equation}\label{cond_igualdadexx}
\frac{1}{\beta}tr(H_j\rho_{\beta}  H_j^*)tr(V_j\rho_{\beta}  V_j^*)=\frac{tr(W_j V_i\rho_W V_i^* W_j^{*})}{tr(V_i\rho_W V_i^*)}
\end{equation}
So we have the following result.
\begin{teo} Let $\mathcal{F}_W$ be a QIFS such that there is a unique attractive invariant measure for the associated Markov operator $\mathcal{V}$. Let $\rho_W$ be the barycenter of such measure and let $\rho_\beta$ be an eigenstate of $\mathcal{L}_H(\rho)$ with eigenvalue $\beta$. Then
\begin{equation}\label{p_var_geralxx} 
h_V(W)+ \sum_{j=1}^k \log \Big( tr(H_j\rho_{\beta}
H_j^*)tr(V_j\rho_{\beta} V_j^*)\Big) tr(W_j \rho_W  W_j^*)
 \leq \log\beta
\end{equation}
and equality holds if and only if for all $i,j$,
\begin{equation}
\frac{1}{\beta}tr(H_j\rho_{\beta}  H_j^*)tr(V_j\rho_{\beta}
V_j^*)=\frac{tr(W_j V_i\rho_W V_i^* W_j^{*})}{tr(V_i\rho_W V_i^*)}
\end{equation}
\end{teo}
In section \ref{sec_analysis4} we make some considerations about
certain cases in which we can reach an equality in
(\ref{p_var_geralxx}).

\qee

For the calculations regarding expression (\ref{soma_1_b}), define
\begin{equation}
r_{jlm}=\frac{1}{\beta}tr(H_j\rho_{\beta}  H_j^*)\frac{(V_j\rho_{\beta}  V_j^*)_{lm}}{(\rho_\beta)_{lm}}
\end{equation}
Then we have $\sum_j r_{jlm}=1$. Let
\begin{equation}
q_{ij}:=tr\Big(\frac{W_j V_i\rho_W V_i^* W_j^{*}}{tr(V_i\rho_W V_i^*)}\Big)
\end{equation}
A calculation similar to the one we have made for (\ref{p_var_geralxx}) gives us
$$h_V(W)+ \sum_{j=1}^k tr(W_j \rho_W  W_j^*) \log tr(H_j\rho_{\beta}  H_j^*)$$
\begin{equation}
+\sum_{j=1}^k tr(W_j \rho_W  W_j^*) \log{\Big(\frac{(V_j\rho_{\beta} V_j^*)_{lm}}{(\rho_\beta)_{lm}}\Big)}
 \leq \log\beta
\end{equation}
and equality holds if and only if for all $i,j,l,m$,
\begin{equation}
\frac{1}{\beta}tr(H_j\rho_{\beta}  H_j^*)\frac{(V_j\rho_{\beta} V_j^*)_{lm}}{(\rho_\beta)_{lm}}=\frac{tr(W_j V_i\rho_W V_i^* W_j^{*})}{tr(V_i\rho_W V_i^*)}
\end{equation}

\qee

\section{Some classic inequality calculations}\label{sec_analysis3}

A natural question is to ask whether the maximum among normalized
$W_i$, $i=1,...,k,$ for the pressure problem associated to a given
potential is realized as the logarithm of the main eigenvalue of a
certain Ruelle operator associated to the potential $H_i$,
$i=1,...,k.$ This problem will be considered in this section and
also in the next one.

We begin by recalling a classic inequality. Consider
\begin{equation}\label{adbas1}
-\sum_{j=1}^kq_j\log{q_j}+\sum_{j=1}^k q_j\log{r_j}\leq 0
\end{equation}
given by lemma \ref{lemalog2}. Let $A$ be a matrix. If $v$ denotes
the left eigenvector of matrix $E^A$ (such that each entry is
$e^{a_{ij}}$), then $vE^A=\beta v$ can be written as
\begin{equation}
\sum_i v_{i}e^{a_{ij}}=\beta v_{j},\sp \forall j
\end{equation}
Define
\begin{equation}
r_{ij}:=\frac{e^{a_{ij}}v_{i}}{\beta v_j}
\end{equation}
So $\sum_i r_{ij}=1$. Let $q_{ij}>0$ such that $\sum_i q_{ij}=1$. By (\ref{adbas1}), we have
\begin{equation}
-\sum_{i=1}^kq_{ij}\log{q_{ij}}+\sum_{i=1}^k q_{ij}\log{\frac{e^{a_{ij}}v_{i}}{\beta v_j}}\leq 0
\end{equation}
That is,
\begin{equation}
-\sum_{i=1}^kq_{ij}\log{q_{ij}}+\sum_{i=1}^k q_{ij}a_{ij}+\sum_{i=1}^k q_{ij}(\log{v_{i}}-\log{v_{j}})\leq \log{\beta}
\end{equation}
Let $Q$ be a matrix with entries $q_{ij}$, let $\pi=(\pi_1,\dots, \pi_k)$ be the stationary vector associated to $Q$. Since $\sum_i q_{ij}=1$, $Q$ is column-stochastic so we write $Q\pi=\pi$. Multiplying the above inequality by $\pi_j$ and summing the $j$ index, we get
\begin{equation}
-\sum_j\pi_j\sum_i q_{ij}\log{q_{ij}}+\sum_j\pi_j\sum_i q_{ij}a_{ij}+\sum_j\pi_j\sum_i q_{ij}(\log{v_{i}}-\log{v_{j}})\leq \log{\beta}
\end{equation}
In coordinates, $Q\pi=\pi$ is $\sum_j q_{ij}\pi_j =\pi_i$, for all $i$. Then
$$-\sum_j\pi_j\sum_i q_{ij}\log{q_{ij}}+\sum_j\pi_j\sum_i q_{ij}a_{ij}$$
\begin{equation}
+\sum_j\pi_j\sum_i q_{ij}\log{v_{i}} -\sum_j\pi_j\sum_i q_{ij}\log{v_{j}}\leq \log{\beta}
\end{equation}
These calculations are well-known and give the following inequality:
\begin{equation}\label{desig_classica}
-\sum_j\pi_j\sum_i q_{ij}\log{q_{ij}}+\sum_j\pi_j\sum_i q_{ij}a_{ij}\leq \log{\beta}
\end{equation}

\begin{defi} We call inequality (\ref{desig_classica}) the {\bf classic inequality} associated to the matrix $A$ with positive entries, and stochastic matrix $Q$.
\end{defi}

\begin{defi} For fixed $k$, and $l,m=1,\dots, k$ we call the inequality
$$h_V(W)+ \sum_{j=1}^k tr(W_j \rho_W  W_j^*) \log tr(H_j\rho_{\beta}  H_j^*)$$
\begin{equation}\label{p_var_denovo}
+\sum_{j=1}^k tr(W_j \rho_W  W_j^*) \log{\Big(\frac{(V_j\rho_{\beta} V_j^*)_{lm}}{(\rho_\beta)_{lm}}\Big)}
 \leq \log\beta,
\end{equation}
the {\bf basic inequality} associated to the potential $H\rho=\sum_i H_i\rho H_i^*$ and to the QIFS determined by $V_i$, $W_i$, $i=1,\dots, k$. Equality holds if for all $i,j,l,m$,
\begin{equation}
\frac{1}{\beta}tr(H_j\rho_{\beta}  H_j^*)\frac{(V_j\rho_{\beta} V_j^*)_{lm}}{(\rho_\beta)_{lm}}=\frac{tr(W_j V_i\rho_W V_i^* W_j^{*})}{tr(V_i\rho_W V_i^*)}
\end{equation}
\end{defi}



\qee

As before $\rho_{\beta}$ is an eigenstate of $\mathcal{L}_H(\rho)$ and $\rho_W$ is the barycenter of the unique attractive, invariant measure for the Markov operator $\mathcal{V}$ associated to the QIFS $\mathcal{F}_W$. Given the classic inequality (\ref{desig_classica}) we want to compare it to the basic inequality (\ref{p_var_denovo}). More precisely, we would like to obtain operators $V_i$ that satisfy the following: given a matrix $A$ with positive entries and a stochastic matrix $Q$, there are  $H_i$ and $W_i$ such that inequality (\ref{p_var_denovo}) becomes inequality (\ref{desig_classica}). We have the following proposition.

\begin{pro}\cite{BLLT}\label{prop_ba_cl}
Define
\begin{equation}
V_1=\left(
\begin{array}{cc}
1 & 0 \\
0 & 0
\end{array}
\right),\sp V_2=\left(
\begin{array}{cc}
0 & 1 \\
0 & 0
\end{array}
\right)
\end{equation}
\begin{equation}
V_3=\left(
\begin{array}{cc}
0 & 0 \\
1 & 0
\end{array}
\right),\sp V_4=\left(
\begin{array}{cc}
0 & 0 \\
0 & 1
\end{array}
\right)
\end{equation}
Let $A=(a_{ij})$ be a matrix with positive entries and $Q=(q_{ij})$ a two-dimensional column-stochastic matrix. Define
\begin{equation}
H_{11}=\left(
\begin{array}{cc}
\sqrt{e^{a_{11}}} & \sqrt{e^{a_{11}}} \\
0 & 0
\end{array}
\right),\sp H_{12}=\left(
\begin{array}{cc}
\sqrt{e^{a_{12}}} & \sqrt{e^{a_{12}}} \\
0 & 0
\end{array}
\right)
\end{equation}
\begin{equation}
H_{21}=\left(
\begin{array}{cc}
0 & 0 \\
\sqrt{e^{a_{21}}} & \sqrt{e^{a_{21}}}
\end{array}
\right),\sp H_{22}=\left(
\begin{array}{cc}
0 & 0 \\
\sqrt{e^{a_{22}}} & \sqrt{e^{a_{22}}}
\end{array}
\right)
\end{equation}
and also
\begin{equation}
W_1=\left(
\begin{array}{cc}
\sqrt{q_{11}} & 0 \\
0 & 0
\end{array}
\right),\sp W_2=\left(
\begin{array}{cc}
0 & \sqrt{q_{12}} \\
0 & 0
\end{array}
\right)
\end{equation}
\begin{equation}
W_3=\left(
\begin{array}{cc}
0 & 0 \\
\sqrt{q_{21}} & 0
\end{array}
\right),\sp W_4=\left(
\begin{array}{cc}
0 & 0 \\
0 & \sqrt{q_{22}}
\end{array}
\right)
\end{equation}
Then the basic inequality associated to $W_i, V_i, H_i$, $i=1,\dots,4$, $l=m=1$ or $l=m=2$, is equivalent to the classic inequality associated to $A$ and $Q$.
\end{pro}

\begin{examp}
Let
$$H_1=\left(
\begin{array}{cc}
2i & 2i \\
0 & 0
\end{array}
\right), \sp H_2=I,\sp H_3=\left(
\begin{array}{cc}
i\sqrt{2} & i\sqrt{2} \\
0 & 0
\end{array}
\right),\sp H_4=I$$
Then
$$H_1^*=\left(
\begin{array}{cc}
-2i & 0 \\
-2i & 0
\end{array}
\right),\sp H_2^*=I,\sp H_3^*=\left(
\begin{array}{cc}
-i\sqrt{2} & 0 \\
-i\sqrt{2} & 0
\end{array}
\right),\sp H_4^*=I$$
If we suppose the $V_i$ are the same as from proposition \ref{prop_ba_cl}, we have that $\rho_\beta$ is diagonal, so
$$tr(H_1\rho_\beta H_1^*)=4 ,\sp tr(H_2\rho_\beta H_2^*)=1,\sp tr(H_3\rho_\beta H_3^*)=2 ,\sp tr(H_4\rho_\beta H_4^*)=1$$
Then $\mathcal{L}_H(\rho)=\beta\rho$ leads us to
$$4\rho_{11}+\rho_{22}=\beta\rho_{11}$$
$$2\rho_{11}+\rho_{22}=\beta\rho_{22}$$
A simples calculation gives
$$\beta=\frac{5+\sqrt{17}}{2}$$
with eigenstate
$$\rho_\beta=\frac{4}{7+\sqrt{17}}\left(
\begin{array}{cc}
\frac{3+\sqrt{17}}{4} & 0 \\
0 & 1
\end{array}
\right)$$
\end{examp}

We want to calculate the $W_i$ which maximize the basic inequality (\ref{p_var_denovo}). Recall that from proposition \ref{prop_ba_cl}, the choice of $V_i$ we made is such that
$$\frac{(V_j\rho_{\beta} V_j^*)_{lm}}{(\rho_\beta)_{lm}}=1,$$
So
\begin{equation}
h_V(W)+ \sum_{j=1}^k tr(W_j \rho_W  W_j^*) \log tr(H_j\rho_{\beta}  H_j^*)\leq \log\beta
\end{equation}
and equality holds if and only if, for all $i,j,l,m$,
\begin{equation}\label{a_cond_pw}
\frac{1}{\beta}tr(H_j\rho_{\beta}  H_j^*)\frac{(V_j\rho_{\beta} V_j^*)_{lm}}{(\rho_\beta)_{lm}}=\frac{tr(W_j V_i\rho_W V_i^* W_j^{*})}{tr(V_i\rho_W V_i^*)}
\end{equation}
Choose, for instance, $l=m=1$. Then condition (\ref{a_cond_pw}) becomes
\begin{equation}
\frac{1}{\beta}tr(H_j\rho_{\beta}  H_j^*)=\frac{tr(W_j V_i\rho_W V_i^* W_j^{*})}{tr(V_i\rho_W V_i^*)}
\end{equation}
To simplify calculations, write $\widehat{W}_i=W_i^*W_i$ and $\widehat{W}_i=(w_{ij}^i)$. Then we get
\begin{equation}
\frac{tr(H_i\rho_\beta H_i^*)}{\beta}=w_{11}^i=w_{22}^i,\sp i=1,\dots, 4
\end{equation}
So we conclude
\begin{equation}
W_i=\frac{1}{\sqrt{\beta}}\left(
\begin{array}{cc}
\sqrt{tr(H_i\rho_\beta H_i^*)} & 0 \\
0 & \sqrt{tr(H_i\rho_\beta H_i^*)}
\end{array}
\right),\sp i=1,\dots, 4
\end{equation}
That is,
\begin{equation}
W_1=\frac{2}{\sqrt{\beta}}I,\sp W_2=\frac{1}{\sqrt{\beta}}I,\sp
W_3=\frac{\sqrt{2}}{\sqrt{\beta}}I,\sp
W_4=\frac{1}{\sqrt{\beta}}I
\end{equation}
Note that
$$\sum_i W_i^*W_i=\frac{4+\sqrt{2}}{\sqrt{\beta}}I\neq I$$
To solve that, we renormalize the potential. Define
\begin{equation}
\tilde{H}_i:=\sqrt{\alpha} H_i
\end{equation}
where
\begin{equation}
\alpha:=\frac{\sqrt{\beta}}{4+\sqrt{2}}
\end{equation}
Then a calculation shows that $\mathcal{L}_{\tilde{H}}(\rho)=\tilde{\beta}\rho$ gives us the same eigenstate as before, that is $\rho_{\tilde{\beta}}=\rho_\beta$. But note that the associated eigenvalue becomes $\tilde{\beta}=\alpha\beta$. Now, note that it is possible to renormalize the $W_i$ in such a way that we obtain $\tilde{W}_i$ with $\sum_i \tilde{W}_i^*\tilde{W}_i=I$, and that these maximize the basic inequality for the $H_i$ initially fixed. In fact, given the renormalized $\tilde{H}_i$, define
\begin{equation}
\tilde{W}_i=\sqrt{\alpha} W_i,\sp i=1,\dots, 4
\end{equation}
Note that $\sum_i \tilde{W}_i^*\tilde{W}_i=I$. Also we obtain
\begin{equation}
h_V(\tilde{W})+ \sum_{j=1}^k tr(\tilde{W}_j \rho_{\tilde{W}}  \tilde{W}_j^*) \log tr(\sqrt{\alpha}H_j\rho_{\beta} \sqrt{\alpha}H_j^*)\leq \log\alpha\beta
\end{equation}
which is equivalent to
\begin{equation}
h_V(\tilde{W})+ \sum_{j=1}^k tr(\tilde{W}_j \rho_{\tilde{W}}  \tilde{W}_j^*) \log (\alpha tr(H_j\rho_{\beta} H_j^*))\leq \log\alpha+\log\beta
\end{equation}
That is
$$
h_V(\tilde{W})+ \sum_{j=1}^k tr(\tilde{W}_j \rho_{\tilde{W}}  \tilde{W}_j^*) \log \alpha
$$
\begin{equation}
+\sum_{j=1}^k tr(\tilde{W}_j \rho_{\tilde{W}}  \tilde{W}_j^*) \log tr(H_j\rho_{\beta} H_j^*)\leq \log\alpha+\log\beta,
\end{equation}
and cancelling $\log\alpha$, we get the same inequality as for the nonrenormalized $H_i$. As we have seen before, such $\tilde{W}_i$ gives us equality. Hence
\begin{equation}
h_V(\tilde{W})+ \sum_{j=1}^k tr(\tilde{W}_j \rho_{\tilde{W}}  \tilde{W}_j^*) \log tr(H_j\rho_{\beta} H_j^*)=\log\beta
\end{equation}
\qee

\section{Remarks on the problem of pressure and quantum mechanics}\label{sec_analysis4}

One of the questions we are interested in is to understand how to formulate a variational principle for pressure in the context of quantum information theory. An appropriate combination of such theories could have as a starting point a relation between the inequality for positive numbers
$$-\sum_i q_i\log q_i+\sum_i q_i\log p_i\leq 0,$$
(seen in certain proofs of the variational principle of pressure),
and the entropy for QIFS we defined before. We have carried out
such a plan and then we have obtained the basic inequality, which
can be written as
\begin{equation}\label{pvar_ver1yy}
h_V(W)+ \sum_{j=1}^k \log \Big( tr(H_j\rho_{\beta}  H_j^*)tr(V_j\rho_{\beta} V_j^*)\Big) tr(W_j \rho_W  W_j^*)
 \leq \log\beta
\end{equation}
where equality holds if and only if for all $i,j$,
\begin{equation}\label{cond_igualdadeyy}
\frac{1}{\beta}tr(H_j\rho_{\beta}  H_j^*)tr(V_j\rho_{\beta}  V_j^*)=\frac{tr(W_j V_i\rho_W V_i^* W_j^{*})}{tr(V_i\rho_W V_i^*)}
\end{equation}
As we have discussed before, it is not clear that given any
dynamics, we can obtain a measure such that we can reach the
maximum value $\log\beta$. Considering  particular cases, we can
suppose, for instance,  that the $V_i$ are unitary. In this way,
we combine in a natural way a problem of classic thermodynamics,
with an evolution which has a quantum character. In this
particular setting, we have for each $i$ that
$V_iV_i^*=V_i^*V_i=I$ and then the basic inequality becomes
\begin{equation}\label{pvar_ver1ii}
h_V(W)+ \sum_{j=1}^k tr(W_j \rho_W  W_j^*)\log tr(H_j\rho_{\beta}  H_j^*)
 \leq \log\beta
\end{equation}
and equality holds if and only if for all $i,j$,
\begin{equation}\label{cond_igualdadeii}
\frac{1}{\beta}tr(H_j\rho_{\beta}  H_j^*)=tr(W_j V_i\rho_W V_i^* W_j^{*})
\end{equation}
We have the following:
\begin{lem}
Given a QIFS with a unitary dynamics (i.e., $V_i$ is unitary for each $i$),
there are $\hat{W}_i$ which maximize (\ref{pvar_ver1yy}), i.e., such that
\begin{equation}\label{pvar_ver1iij}
h_V(\hat{W})+ \sum_{j=1}^k tr(\hat{W}_j \rho_{\hat{W}}  \hat{W}_j^*)\log tr(H_j\rho_{\beta}  H_j^*)
= \log\beta
\end{equation}
\end{lem}

The above lemma also holds for the basic inequality in coordinates, given by (\ref{p_var_denovo}). Also, it is immediate to obtain a similar version of the above lemma for any QIFS such that the $V_i$ are multiples of the identity, and also for QIFS such that $\rho_W$ fixes each branch of the QIFS, that is, satisfying
$$\frac{V_i\rho_W V_i^*}{tr(V_i\rho_W V_i^*)}=\rho_W$$


\begin{thebibliography}{99}

\bibitem{BLLT} Baraviera, A., Lardizabal, C. F., Lopes, A. O.,
Terra Cunha, M. A Thermodynamic Formalism for density matrices in
Quantum Information. Applied Mathematics Research Express, Vol. 2010, No.1, pp. 63-118.

\bibitem{BLLT2} Baraviera, A., Lardizabal, C. F., Lopes, A. O.,
Terra Cunha, M. Quantum Stochastic Processes, Quantum Iterated
Function Systems and Entropy. To appear on S\~ao Paulo Journal of Mathematical Sciences (2010).

\bibitem{benatti} Benatti, F. Dynamics, Information and Complexity in Quantum Systems.    Springer, 2009.

\bibitem{bcs} Benenti, G., Casati, G. Strini, G.
Principles of Quantum Computation
and Information, Vol I and II,  World scientific, 2007

\bibitem{bengts} Bengtsson, I., \.Zyczkowski, K. Geometry of Quantum States. Cambridge University Press, 2006.

\bibitem{busch} Busch, P., Ruch, E. The measure cone: irreversibility as a geometrical phenomenon, Int. J. Quantum Chemistry. 41, 163-185, 1992.

\bibitem{GL} Castro, G., Lopes, A. O.
KMS States, Entropy and a Variational Principle for Pressure, to
appear in Real Analysis Exchange (2009).


\bibitem{GZ}
Gardiner. C. W., Zoller, P. Quantum Noise, Springer Verlag, 2004.

\bibitem{gray} Gray, R. M. Entropy and information theory. Springer-Verlag, New York, 1990.

\bibitem{gosson} de Gosson, M. Symplectic geometry and quantum mechanics. Birkhauser, 2006.

\bibitem{sigal} Gustafson, S., Sigal, I. Mathematical Concepts of Quantum Mechanics. Springer-Verlag, 2003.

\bibitem{hastings} Hastings, M. B. A counterexample to additivity of minimum output entropy. arXiv:0809.3972v3 [quant-ph], 2008.

\bibitem{hayashi1} Hayashi, M. Capacity with energy constraint in coherent state channel.    arXiv:0904.0307v1 [quant-ph], 2009.

\bibitem{hayashi2} Hayashi, M., Nagaoka, H. General formulas for capacity of classical-quantum channels. IEEE Transactions on Information Theory, 7, v. 49, 2003.

\bibitem{jordan} Jordan, T. Affine maps of density matrices. Physical Review A, 71, 034101, 2005.

\bibitem{lar} Lardizabal, C. F. Processos Estoc\'asticos Qu\^anticos, Tese de doutorado Prog. Posgrad. Mat. UFRGS - to appear
(2010).

\bibitem{lasota} Lasota, A., Mackey, M. Chaos, fractals and noise. Springer-Verlag, New York, 1994.

\bibitem{lopes_elismar} Lopes, A. O., Oliveira, E. Entropy and variational principles for holonomic probabilities of IFS. Discrete and Continuous Dynamical Systems Vol. 23, N, 3, Series A. 937-955, 2009.

\bibitem{lop1} Lopes, A. O. Entropy and Large Deviation. NonLinearity Vol. 3, N. 2, 527-546, 1990.

\bibitem{lopes} Lopes, A. O. An analogy of the charge distribution on Julia sets with the Brownian motion. J. Math. Phys. 30 (9), 1989.

\bibitem{lopescra} Lopes, A. 0., Craizer, M. The capacity-cost function of a hard-constrained channel. Int. Journal of Appl. Math. Vol 2, N 10 pp 1165-1180 (2000).

\bibitem{lozinski} Lozinski, A., \.Zyczkowski, K., S\l omczy\'{n}ski, W. Quantum iterated function systems, Physical Review E, Volume 68, 04610, 2003.

\bibitem{Man} Ma\~n\'e, R. Ergodic Theory, Springer Verlag, 1986.

\bibitem{nich} Nielsen, M., Chuang, I. Quantum computation and quantum information. Cambridge University Press, 2000.

\bibitem{par} Parry, W., Pollicott, M. Zeta Functions and the Periodic
Orbit Structure of Hyperbolic Dynamics. Soci\'et\'e Math\'ematique
de France. 187-188, Ast\'erisque, 1990.

\bibitem{R1} Rieffel, E., Polak, W.
An Introduction to Quantum Computing for
Non-Physicists, ACM Computing Surveys,   Vol. 32,   Issue: 3,  p. 300-335,   2000.

\bibitem{shor} Shor, P. W. Equivalence of additivity question in quantum information theory. Comm. Math. Phys. 246, 453-472 (2004).

\bibitem{slom} S\l omczy\'{n}ski, W., \.Zyczkowski, K. Quantum Chaos: an entropy approach. J. Math. Physics, 32 (1), 1994, p. 5674-5700.

\bibitem{wsbook} S\l omczy\'{n}ski, W., Dynamical Entropy, Markov Operators and Iterated Function Systems.  Jagiellonian University Press, 2003.

\bibitem{Sp} Spitzer, F.
A Variational characterization of finite Markov chains. The Annals
of Mathematical Statistics. (43): N.1  303-307, 1972.

\bibitem{Sr} Srinivas, M. D. Foundations of a quantum probability theory
Journal of Math. Phys., Vol. 16, No. 8, 1975.

\bibitem{winkler} Winkler, G., Choquet Order and Simplices. Lecture notes in Mathematics 1145. Springer-Verlag, Berlin, 1985.


\end{thebibliography}
\end{document}